\documentclass[a4paper,12pt]{article}

\pdfoutput=1

\usepackage{amsmath}
\usepackage{amssymb}
\usepackage{graphicx}
\usepackage{hyperref}
\usepackage{cite}
\usepackage{color}

\makeatletter
\@addtoreset{equation}{section}
\renewcommand{\theequation}{\thesection.\@arabic\c@equation}
\makeatother

\begin{document}

\begin{titlepage}

\vspace*{-15mm}   
\baselineskip 10pt   
\begin{flushright}   
\begin{tabular}{r}    
\end{tabular}   
\end{flushright}   
\baselineskip 24pt   
\vglue 10mm   

\begin{center}
{\Large\bf
 Universal structure of islands\\ in evaporating black holes
}

\vspace{8mm}   

\baselineskip 18pt   

\renewcommand{\thefootnote}{\fnsymbol{footnote}}

Yoshinori Matsuo\footnote[2]{ymatsuo@phys.kindai.ac.jp} 

\renewcommand{\thefootnote}{\arabic{footnote}}
 
\vspace{5mm}   

{\it  
 Department of Physics, Kindai University, \\Higashi-Osaka, Osaka 577-8502, Japan
}
  
\vspace{10mm}   

\end{center}

\begin{abstract}

The entanglement entropy of the Hawking radiation contains 
contributions from a region inside the black hole, which is called islands, 
implying that the Hawking radiation contains the information of islands. 
The boundary of the island is given by the quantum extremal surface, 
whose position is determined so that the entanglement entropy is extremized. 
In many cases of stationary black holes and a few cases of evaporating black holes, 
it was already confirmed that the quantum extremal surface is located 
outside the horizon for stationary black holes and 
is inside the horizon for evaporating black holes. 
In this paper, we calculate islands in general black holes and show that 
the island extends to the outside of the horizon for stationary black holes 
but is hidden inside the horizon for evaporating black holes 
independent of details of the black hole. 

\end{abstract}

\baselineskip 18pt   

\end{titlepage}

\newpage

\baselineskip 18pt

\tableofcontents


\section{Introduction}\label{sec:Intro}

One of the most important discovery in the recent progress in black hole physics would be islands 
\cite{Penington:2019npb,Almheiri:2019psf,Almheiri:2019hni,Almheiri:2019yqk}.%
The Hawking radiation \cite{Hawking:1974sw} arises simply by the time evolution of the vacuum state 
and hence naively contains no information about the matter which entered the black hole. 
In order to preserve unitarity in black hole evaporation, 
the Hawking radiation must carry the information of the matter inside the black hole. 
This inconsistency is the notorious information loss paradox in black hole physics \cite{Hawking:1976ra}. 
The prescription of islands is proposed to reproduce the Page curve \cite{Page:1993wv,Page:2013dx}. 
The entanglement entropy of the Hawking radiation follows the Page curve 
if the Hawking radiation contains the information of matter inside the black hole. 
How to reproduce the Page curve is an aspect of the information loss problem. 

The formula of islands is first proposed by using 
the prescription of the quantum extremal surface 
\cite{Faulkner:2013ana,Engelhardt:2014gca,Dong:2016hjy,Dong:2017xht}, 
which is a generalization of the Ryu-Takayanagi formula to calculate 
the entanglement entropy in holography \cite{Ryu:2006bv,Hubeny:2007xt,Lewkowycz:2013nqa}. 
Soon after the first proposal, it was pointed out that 
the island formula can be reproduced by using the replica trick 
\cite{Callan:1994py,Holzhey:1994we,Calabrese:2004eu,Casini:2005rm,Calabrese:2009qy} in theories 
with gravity \cite{Penington:2019kki,Almheiri:2019qdq}, 
and hence, the formula can be applied for non-holographic setups. 
The formula states that the entanglement entropy of the Hawking radiation at late times 
effectively contains the information of a region inside the black hole called islands $I$. 
To be more precise, the entanglement entropy of the Hawking radiation is identified 
with the entanglement entropy of fields in a region $R$ which is placed outside the black hole. 
Then, the entanglement entropy of the region $R$ is effectively given by 
the entanglement entropy of $R \cup I$. 
(See \cite{Almheiri:2020cfm} for a review and 
\cite{Akers:2019nfi,Chen:2019uhq,Almheiri:2019psy,Chen:2019iro,Akers:2019lzs,Liu:2020gnp,Marolf:2020xie,Balasubramanian:2020hfs,Bhattacharya:2020ymw,Verlinde:2020upt,Chen:2020wiq,Gautason:2020tmk,Anegawa:2020ezn,Hashimoto:2020cas,Sully:2020pza,Hartman:2020swn,Hollowood:2020cou,Krishnan:2020oun,Alishahiha:2020qza,Banks:2020zrt,Geng:2020qvw,Chen:2020uac,Chandrasekaran:2020qtn,Li:2020ceg,Bak:2020enw,Bousso:2020kmy,Anous:2020lka,Dong:2020uxp,Krishnan:2020fer,Hollowood:2020kvk,Engelhardt:2020qpv,Karlsson:2020uga,Chen:2020jvn,Chen:2020tes,Hartman:2020khs,Liu:2020jsv,Murdia:2020iac,Akers:2020pmf,Balasubramanian:2020xqf,Balasubramanian:2020coy,Sybesma:2020fxg,Stanford:2020wkf,Chen:2020hmv,Ling:2020laa,Marolf:2020rpm,Harlow:2020bee,Akal:2020ujg,Hernandez:2020nem,Chen:2020ojn,Matsuo:2020ypv,Goto:2020wnk,Hsin:2020mfa,Akal:2020twv,Colin-Ellerin:2020mva,KumarBasak:2020ams,Geng:2020fxl,Karananas:2020fwx,Wang:2021woy,Marolf:2021kjc,Bousso:2021sji,Geng:2021wcq,Ghosh:2021axl,Wang:2021mqq,Uhlemann:2021nhu,Kawabata:2021vyo,Akal:2021foz,Omiya:2021olc,Geng:2021hlu,Ahn:2021chg,Stanford:2021bhl,Balasubramanian:2021xcm,Akers:2021lms,Miyaji:2021lcq,Azarnia:2021uch,He:2021mst,Dong:2021oad,Shaghoulian:2021cef,Matsuo:2021mmi,Uhlemann:2021itz,Omidi:2021opl,Bhattacharya:2021nqj,Yu:2021rfg,Engelhardt:2022qts,Akers:2022max,Suzuki:2022xwv,Gan:2022jay,Rolph:2022csa,Bousso:2022tdb,Akers:2022qdl,Balasubramanian:2022gmo,Wu:2023uyb,Jeong:2023lkc,Afrasiar:2023nir,Matsuo:2023cmb,Bousso:2023kdj} 
for related works.) 

The location of islands $I$ depends on whether the black hole is 
stationary or dynamical \cite{Penington:2019npb,Gautason:2020tmk,Matsuo:2020ypv}. 
In the case of stationary black holes, or equivalently, in eternal black hole geometries, 
the island extends to slightly outside the horizon. 
In the case of dynamical black holes, which are formed by a gravitational collapse of matter, 
the island is located inside the horizon. 
The difference comes from several different structures in two cases, 
but the most important factor would be vacuum states. 
In the case of stationary black holes, the vacuum state of matter fields 
must be the Hartle-Hawking vacuum \cite{Hartle:1976tp} to avoid the divergence on the horizon. 
The Hartle-Hawking vacuum is also consistent with stationary nature of the black hole 
as it contains the incoming radiation 
which is balanced with the (outgoing) Hawking radiation. 
In contrast, dynamical black hole geometries have no past horizon, and hence, 
the regularity condition on the past horizon does not need to be satisfied. 
Instead, the initial state before the collapse should be the vacuum state in flat spacetime. 
Thus, the vacuum state around dynamical black holes is the Unruh vacuum \cite{Unruh:1976db}. 
The Unruh vacuum has Hawking radiation but no incoming radiation in the asymptotic region. 
Then, the black hole only loses the energy and eventually evaporates. 
The difference of the vacuum states --- the Hartle-Hawking vacuum or Unruh vacuum --- 
is most important for the different location of the island in two cases. 

Another important effect is that the position of the horizon changes with time in dynamical cases 
because the black hole is evaporating. 
The position of the quantum extremal surface, which is the boundary of the island, 
is determined so that the entanglement entropy is extremized. 
In the case of the evaporating Schwarzschild black hole, 
the entanglement entropy of Hawking radiation is expressed by using the s-wave approximation 
as \cite{Penington:2019npb,Matsuo:2020ypv,Matsuo:2023cmb}
\begin{equation}
 S(R) 
 \simeq 
 \frac{\pi r_h^2(v)}{G_N} 
 + \frac{c v}{24 r_h(v)} + \cdots \ , 
\label{Schwarz}
\end{equation}
where $r_h$ is the Schwarzschild radius and $v$ is the advanced time. 
The first term comes from the gravity part, which is given by the Bekenstein-Hawking entropy, 
and the second term is contribution from the matter part, 
or equivalently, effects of the Unruh vacuum. 
Because of the effect of the second term in \eqref{Schwarz}, 
the quantum extremal surface is located inside the horizon. 
However, the Schwarzschild radius $r_h(v)$ decreases with time 
due to the emission of the Hawking radiation as 
\begin{equation}
 \dot r_h(v) = - \frac{c\,G_N}{96\pi r_h^2(v)} \ , 
\end{equation}
and then, the entanglement entropy is expressed as 
\begin{equation}
 S(R) 
 \simeq 
 \frac{\pi r_h^2(v=0)}{G_N} - \frac{c v}{48 r_h(v)} 
 + \frac{c v}{24 r_h(v)} + \cdots 
 \ . 
 \label{Schwarz1}
\end{equation}
Thus, a half of the second term in \eqref{Schwarz} 
is canceled by the decrease of the area term. 
The quantum extremal surface is located inside the horizon 
because the second term in \eqref{Schwarz1} is smaller than the third term. 
If the horizon radius decreased much faster or if the effect of the Unruh vacuum is much smaller, 
the quantum extremal surface would be outside the horizon. 

In this paper, we consider islands in general black hole geometries 
to see explicitly that the structure above of islands 
is universal and independent of details of the black hole. 
We first show that the island extends to outside the horizon 
independent of details of the geometry in the case of stationary black holes. 
If only the gravity part is considered, the classical extremal surface is 
located at the bifurcation surface of the horizon. 
Due to effects of the matter in the bulk outside the horizon, 
the extremal surface is moved outward from the horizon. 
Thus, two boundaries of the island, which are the quantum extremal surfaces, 
are separated and located in each of two exteriors of the horizon. 
As the effect of the matter is much smaller than the effect of gravity, 
the quantum extremal surfaces stay very near the horizon. 
Thus, for any black hole, the entanglement entropy of the Hawking radiation 
is always given approximately by the black hole entropy. 

Secondly, we show that the island is located inside the horizon 
in the case of evaporating black holes. 
The black hole is evaporating by emitting the Hawking radiation, 
or equivalently, due to effects of the Unruh vacuum. 
Thus, the effect of the decreasing horizon radius is 
related to the effect of the Unruh vacuum through thermodynamic relations. 
In determining the position of the quantum extremal surface, 
the effect of decreasing horizon radius is always smaller than 
the effect of the Unruh vacuum, and hence, the quantum extremal surface is 
inside the horizon independent of details of the geometry. 

This paper is organized as follows. 
In Sec.~\ref{sec:Island}, we briefly review the formula of islands. 
In Sec.~\ref{sec:Eternal}, we study islands in general stationary black holes. 
In Sec.~\ref{sec:Evaporating}, we consider islands in evaporating black holes. 
Sec.~\ref{sec:Conclusion} is devoted to the conclusion and discussions.


\section{Island rule}\label{sec:Island}

In this section, we briefly review the island rule 
for calculating the entanglement entropy of the Hawking radiation. 
The entanglement entropy of the Hawking radiation is identified with 
that of a region outside the event horizon, which is referred to as $R$. 
The island rule states that the entanglement entropy of a region $R$ is given by 
\begin{equation}
 S(R) 
 = 
 \min\left\{\mathrm{ext}\left[
 \sum_{\Sigma=\partial R,\partial I} S_\text{grav}(\Sigma) + S_\text{matter}(R\cup I)
 \right]\right\} \ , 
 \label{S(R)}
\end{equation}
where $S_\text{grav}$ and $S_\text{matter}$ 
are the gravity part and matter part of the entanglement entropy, respectively. 
The entanglement entropy effectively given by that of the union of the region $R$ and islands $I$, 
and the gravity part has contributions from all boundaries of $R\cup I$. 
The extremum should be taken for the configuration (position) of islands, and then, 
the minimum should be chosen in extrema for configurations with arbitrary numbers of islands 
--- including the case without islands. 
For the Einstein gravity, whose action $\mathcal I_\text{grav}$ is given by 
\begin{equation}
 \mathcal I_\text{grav} 
 = 
 \frac{1}{16\pi G_N} \int_{\mathcal M} d^Dx \sqrt{-g}\,R 
 + \frac{1}{8\pi G_N} \int_{\partial\mathcal M} d^{D-1}x \sqrt{-h}\,K \ , 
 \label{I_grav}
\end{equation}
where $R$ is the scalar curvature and $K$ is (the trace of) the extrinsic curvature, 
the gravity part of the entanglement entropy is given by 
the area terms \cite{Bombelli:1986rw,Srednicki:1993im} 
\begin{equation}
 S_\text{grav}(\Sigma) = 
 \frac{\mathrm{Area}(\Sigma)}{4G_N} \ ,  
 \label{S-ein}
\end{equation}
which is nothing but the Bekenstein-Hawking entropy when the surface $\Sigma$ is the horizon. 
We will ignore the contribution from the boundary of $R$ in the gravity part, 
namely $S_\text{grav}(\partial R)$, 
as it is independent of the state of the Hawking radiation. 
Only the boundaries of islands make 
important contributions in the gravity part of the entanglement entropy of the Hawking radiation. 

The formula \eqref{S(R)} can be derived by using the replica trick, 
\begin{equation}
 S = \lim_{n\to 1} \frac{1}{1-n}\log \frac{Z_n}{Z_1^n} \ , 
 \label{replica}
\end{equation}
where $Z_n$ is the partition function for the $n$-sheeted spacetime. 
We insert a branch cut(s) on the region $R$ by hand. 
The path integral for the replica spacetime in a theory with gravity 
contains configurations with replica wormholes which connect different copies of the spacetime. 
Such replica wormholes have the same effect to branch cuts in the replica spacetime 
and become islands in the formula \eqref{S(R)}. 

We assume that the gravity part can be treated classically%
\footnote{%
This implies that the coefficient of the gravity part of the action is sufficiently large. 
In the case of the Einstein gravity, for example, 
the Newton constant $G_N$ must be sufficiently small compared with the typical length scale of the system. 
}
and take only the most dominant saddle point in the path integral. 
The difference between the replica spacetime and ($n$-copies of) the original spacetime 
can be pushed to the conical singularity at the branch points. 
For the Einstein gravity \eqref{I_grav}, 
the conical singularity along boundaries of $R\cup I$ 
gives the area terms. 
For more general gravity theories, 
the gravity part of the entanglement entropy is given by 
the Wald entropy \cite{Wald:1993nt,Iyer:1994ys}
with additional anomaly terms as \cite{Dong:2013qoa,Camps:2013zua}
\begin{align}
 S_\text{grav} 
&= 
 S_\text{Wald} + S_\text{anom} 
 \ , 
\label{S_grav}
\\
 S_\text{Wald} 
 &= 
 - 2\pi \int_{\Sigma} d^{D-2} x \sqrt{\sigma} 
 \frac{\partial\mathcal L}{\partial R_{\mu\nu\rho\sigma}} \epsilon_{\mu\rho}\epsilon_{\nu \sigma} \ , 
\label{S_Wald}
\\
 S_\text{anom} 
 &= 
 2\pi \int_{\Sigma} d^{D-2} x \sqrt{\sigma}  
 \sum_\alpha 
  \left(
   \frac{\partial^2\mathcal L}{\partial R_{\mu_1\nu_1\rho_1\sigma_1}\partial R_{\mu_2\nu_2\rho_2\sigma_2}} 
  \right)_\alpha 
  \frac{2 K_{\lambda_1\rho_1\sigma_1}K_{\lambda_2\rho_2\sigma_2}}{q_\alpha+1} 
  P_{\lambda_1\rho_1\sigma_1\lambda_2\rho_2\sigma_2} \ , 
\label{S_anom}
\end{align}
where $\mathcal L$ is the lagrangian density of the gravity theory, 
$\sigma_{\mu\nu}$ is the induced metric on the codimension-2 surface $\Sigma$, 
$K_{\lambda\mu\nu} = n_\lambda^{(i)} K_{(i)\mu\nu}$ is the (two independent) extrinsic curvature on $\Sigma$, 
$\epsilon_{\mu\nu}$ is the volume form of the two-dimensional space transverse to $\Sigma$, 
and $P_{\lambda_1\rho_1\sigma_1\lambda_2\rho_2\sigma_2}$ is a projection 
to $\mu_1=\nu_1=\lambda_1\neq \mu_2=\nu_2=\lambda_2$ in this two-dimensional space. 
In terms of $n_{\mu}^{(i)}$, which are two orthogonal unit normal vectors to $\Sigma$, 
we have 
\begin{align}
 \sigma_{\mu\nu} 
 &= 
 g_{\mu\nu} - n_{\mu\nu} \ , 
 \\
 n_{\mu\nu} 
 &= 
 \sum_{i=1}^2 n_\mu^{(i)} n_\nu^{(i)} \ , 
 \\
 \epsilon_{\mu\nu} 
 &= 
 n_\mu^{(i)} n_\nu^{(j)} \epsilon_{ij} \ , 
 \\
 P_{\lambda_1\rho_1\sigma_1\lambda_2\rho_2\sigma_2} 
 &= 
 \left(n_{\mu_1\mu_2} n_{\nu_1\nu_2} - \epsilon_{\mu_1\mu_2} \epsilon_{\nu_1\nu_2} \right) 
 n^{\lambda_1\lambda_2}
 + \left(n_{\mu_1\mu_2} \epsilon_{\nu_1\nu_2} + \epsilon_{\mu_1\mu_2} n_{\nu_1\nu_2} \right) 
 \epsilon^{\lambda_1\lambda_2} \ , 
\end{align}
where $\epsilon_{ij}$ (with Latin indices) is the Levi-Civita symbol. 
The second derivative of the lagrangian density is expanded around the conical singularity as 
\begin{equation}
 \frac{\partial^2\mathcal L}{\partial R_{\mu_1\nu_1\rho_1\sigma_1}\partial R_{\mu_2\nu_2\rho_2\sigma_2}} 
 = 
 \sum_\alpha 
  \left(
   \frac{\partial^2\mathcal L}{\partial R_{\mu_1\nu_1\rho_1\sigma_1}\partial R_{\mu_2\nu_2\rho_2\sigma_2}} 
  \right)_\alpha \rho^{2\left(1-\frac{1}{n}\right)q_\alpha} + \cdots \ , 
\label{anom-exp}
\end{equation}
where $\rho$ is the distance from the conical singularity, 
and we have taken only the terms which do not vanish at $\rho=0$ in $n\to 1$. 
Then, the coefficients of the second term in \eqref{S_anom} and the parameters $q_\alpha$ 
can be read off from the expression \eqref{anom-exp}. 

The matter part of the entanglement entropy can be separated into 
divergent terms and finite part, which correspond to those in the effective matter action. 
After integrating out the matter fields, 
the partition function of the matter part can be expressed as a functional of the metric. 
Before the renormalization, the partition function contains divergent terms, 
which have local expressions as they come from physics near the cut-off scale. 
For example, the one-loop effective partition function of free massless fields in four dimensions 
can be expressed as 
\begin{align}
 \log Z 
 &= 
 \frac{1}{\epsilon^4}\int d^4x \sqrt{-g}\ c_0 
 + \frac{1}{\epsilon^2} \int d^4x \sqrt{-g}\ c_1 R 
\notag\\
&\quad
 + \log \epsilon \int d^4x \sqrt{-g} 
  \left(
   c_2 R^{\mu\nu\rho\sigma} R_{\mu\nu\rho\sigma}
   + c_3 R^{\mu\nu} R_{\mu\nu}
   + c_4 R^2 
  \right)
\notag\\
&\quad
 + \text{(Regular terms)} \ , 
\label{1loop}
\end{align}
where $\epsilon$ is the UV cut-off. 
These UV divergent terms take the same form to terms in the gravitational action, 
and the divergence should be canceled by introducing appropriate counter terms, 
or equivalently, the sum of the bare gravity action and the UV divergent terms 
gives a finite physical gravity action \cite{Susskind:1994sm}. 
For non-gravitating theories, by applying the replica trick \eqref{replica} 
for the partition function \eqref{1loop}, 
divergent terms in the effective action above yield divergent terms in the entanglement entropy. 
For example, the second term in \eqref{1loop} has the same form to the Einstein-Hilbert action 
and gives the area term of the entanglement entropy. 
For theories with gravity, these divergent terms are absorbed by 
the gravitational action as the renormalization of coupling constants, 
and then, their contributions are included in the gravity part $S_\text{grav}$ in \eqref{S(R)}. 

The other terms of the matter part give $S_\text{matter}$. 
For $S_\text{matter}$, 
the finite part of the partition function on the replica spacetime 
can be calculated by inserting twist operators at the branch points. 
It is very difficult to calculate the correlation function of twist operators in general theories 
as the explicit form of the twist operator is not known in most theories, 
while they can be calculated explicitly in several models in two dimensions. 
For the massless free fermions in two dimensions, 
the entanglement entropy is evaluated explicitly as 
\begin{equation}
 S 
 = 
 \frac{c}{3} \sum_{i,j}\log\left|x_i-x'_j\right| 
 - \frac{c}{6} \sum_{i,j}\log\left|x_i-x_j\right| 
 - \frac{c}{6} \sum_{i,j}\log\left|x'_i-x'_j\right| \ ,  
 \label{Smatter0}
\end{equation}
where $c$ is the central charge, and $(x_i, x_i')$ are two branch points of $i$-th branch cut. 
This expression has divergence which comes from the interaction of each twist operator with itself. 
The divergence should be regularized by introducing a UV cut-off, 
which should be given in terms of a proper physical scale, $\epsilon\sim \Delta s$. 
However, the distance $|x-y|$ in the expression \eqref{Smatter0} is not a proper distance, 
but defined by the distance up to the conformal factor in a coordinate system in the conformal gauge, 
\begin{equation}
 ds^2 = e^{2\rho} \eta_{\mu\nu} dx^\mu dx^\nu \ . 
\end{equation}
The divergent terms in \eqref{Smatter0} is regularized 
by using a physical cut-off scale $\epsilon$ as 
\begin{equation}
 \log\left|\Delta x\right| = -\rho + \log \epsilon \ . 
\end{equation}
Then, the entanglement entropy is expressed as 
\begin{align}
 S 
 &= 
 \frac{c}{3} \sum_{i,j}\log\left|x_i-x'_j\right| 
 - \frac{c}{3} \sum_{i<j}\log\left|x_i-x_j\right| 
 - \frac{c}{3} \sum_{i<j}\log\left|x'_i-x'_j\right| 
\notag\\&\quad 
 + \frac{c}{6} \sum_i \rho(x_i) 
 + \frac{c}{6} \sum_i \rho(x'_i) 
 - \frac{c\,k}{3} \log \epsilon \ , 
\label{Smatter}
\end{align}
where $k$ is the number of branch cuts. 
The last term in \eqref{Smatter} is the divergent term 
and is absorbed by the renormalization with the gravity part. 
The entanglement entropy calculated by using the replica trick is 
basically for the vacuum state, unless the euclidean time is compactified. 
The vacuum state depends on the coordinate system --- 
a vacuum state in a coordinate system is not the vacuum state in another coordinate system. 
The expression \eqref{Smatter} is not given in a covariant form and depends on coordinates. 
It gives the entanglement entropy in the vacuum state 
in the coordinate system which is used in the expression. 
For more details, see Appendix. 

The entanglement entropy for more general conformal field theories 
have additional terms to the expression above, 
but eq.~\eqref{Smatter} gives a universal exact expression of the entanglement entropy 
for arbitrary two-dimensional conformal field theories if the region is a connected interval. 
As we will discuss in the subsequent sections, 
a formula for a connected interval is sufficient 
to evaluate the entanglement entropy of the Hawking radiation. 

In this paper, we use the s-wave approximation to approximate fields around the black hole, 
of which the Hawking radiation consists, by two-dimensional massless free fields. 
We assume that there are no massive fields and ignore 
the higher Kaluza-Klein modes, which has effective Kaluza-Klein mass 
after the dimensional reduction to two dimensions. 
Although the formula \eqref{Smatter} does not give the exact expression of the entanglement entropy 
for two or more connected intervals with general two-dimensional matter fields, 
we effectively need to calculate only the entanglement entropy of a connected region, as we will see later. 
Thus, the matter part of the entanglement entropy $S_\text{matter}$ 
is approximated by using the expression \eqref{Smatter}.


\section{Islands in eternal black holes}\label{sec:Eternal}

In this section, we consider the entanglement entropy of 
the Hawking radiation around a stationary black hole. 
We first briefly review the structure of the geometry near the horizon 
of stationary black holes in Sec.~\ref{ssec:NHstat}. 
Then, we calculate the entanglement entropy of the Hawking radiation 
in configurations without islands in Sec.~\ref{ssec:NoIslandStat} 
and with an island in Sec.~\ref{ssec:IslandStat}. 
In these calculations, we focus on the geometry whose angular directions are 
factorized except for the overall scaling factor. 
We use the s-wave approximation, and consider two-dimensional massless fields. 
In Sec.~\ref{ssec:General}, we consider further generalization. 
In Sec.~\ref{sssec:GeneralMatter}, we calculate the entanglement entropy for the general matter. 
Then, we discuss the entanglement entropy in general black holes 
including the rotating black holes in Sec.~\ref{sssec:GeneralBH}.


\subsection{Near horizon geometry of stationary black holes}\label{ssec:NHstat}

Stationary spacetimes possess a Killing vector, and 
the event horizon of stationary black holes coincides with the Killing horizon. 
We define the advanced time $v$ so that the displacement along a null generator 
is given by $dx^\mu = \xi^\mu dv$, 
where $\xi^\mu$ is the Killing vector tangent to the null generator, 
or equivalently, 
\begin{equation}
 \xi = \partial_v \ . 
 \label{Killing0}
\end{equation}
However, the vector \eqref{Killing0} does not satisfy the Killing equation 
as it is the Killing vector only on the Killing horizon. 
The Killing vector in the neighborhood of the horizon is expressed as 
\begin{equation}
 \xi = \partial_v + \xi^U \partial_U \ , 
\end{equation}
where $U$ parameterizes the displacement from the Killing horizon, 
and $\xi^U = 0$ on the Killing horizon $U=0$. 
Note that $\xi$ is not tangent to a geodesic away from the Killing horizon. 
The Killing equation can be satisfied if the metric has the isometry generated by $\xi$, 
and $\xi^U$ is determined by solving the Killing equation on the stationary spacetime. 

The metric in the vicinity of the horizon is expressed as 
\begin{equation}
 ds^2 = - C(U,v) dUdv + \sigma_{ij} d \theta^i d \theta^j \ , 
 \label{NHmetric}
\end{equation}
where $\theta^i$ is the spatial coordinates on the horizon, 
and $\sigma_{ij}$ is the induced metric on the time slice. 
The Killing equation for $\xi$ implies that 
\begin{equation}
 C(U=0,v) = e^{\kappa v} \ , 
\label{gUv}
\end{equation}
up to an overall constant, and 
the other components of the metric is constant along the null generators of the horizon, 
\begin{equation}
 \xi^\mu \partial_\mu \sigma_{ij} = 0 \ . 
\end{equation}

The constant $\kappa$ is related to the Killing vector, 
or equivalently, to the advanced time $v$. 
%
As it can be seen in the expression \eqref{NHmetric}, 
the advanced time $v$ is not an affine parameter. 
By solving the Killing equation, we obtain 
\begin{equation}
 \xi^U = - \kappa U + \mathcal O(U^2) \ , 
\end{equation}
and then, the geodesic equation for the Killing vector is expressed 
by in terms of $\kappa$ in \eqref{gUv} as 
\begin{equation}
 \xi^\nu \nabla_\nu \xi^\mu = \kappa \xi^\mu \ .  
 \label{geodesic}
\end{equation}
Therefore, $\kappa$ is nothing but the surface gravity of the black hole. 
The Killing vector is expressed by using an affine parameter $V$ as 
\begin{equation}
 \xi = \kappa V \partial_V \ ,  
\end{equation}
on the Killing horizon. 
This implies that the Killing horizon has the bifurcation surface --- 
a fixed point of the isometry, on which $\xi=0$, at $V=0$. 
We can choose the retarded time so that 
the metric $C(U,v)$ is independent of $U$ at $V=0$, 
and then, the metric is expressed as 
\begin{equation}
 ds^2 = - dU\,dV + \sigma_{ij} d \theta^i d \theta^j \ . 
 \label{NHmetric1}
\end{equation}
The surface $V=0$ is also a Killing horizon and 
the Killing vector near the horizon is expressed as  
\begin{equation}
 \xi = \kappa V \partial_V - \kappa U \partial_U \ , 
\label{Killing}
\end{equation}
Thus, stationary black holes have a bifurcate horizon at $UV=0$ ($U=0$ or $V=0$), 
and the metric near the horizon is given by \eqref{NHmetric1}. 

The metric of general black holes can be expressed as 
\begin{equation}
 ds^2 
 = 
 - e^{2\rho} dU\,dV 
 + \sigma_{ij} 
 \left(d \theta^i + A^i_\alpha dx^\alpha\right) 
 \left(d \theta^j + A^j_\beta dx^\beta\right) \ . 
 \label{STmetric}
\end{equation}
where $x^\alpha = (U,V)$ is the (null) coordinates in temporal and radial directions. 
Since \eqref{STmetric} is the most general expression of the metric, 
the coordinates can be chosen to agree with \eqref{NHmetric1} on the horizon. 
Then, we have 
\begin{equation}
 \rho = A^i_\alpha = 0 \ , 
 \label{NHgauge}
\end{equation}
on the bifurcate horizon, $UV=0$. 
The off-diagonal components $A^i_\alpha$ represent 
stationary motion of the black hole, such as a rotation. 
The gauge condition \eqref{NHgauge} is for the comoving frame of the horizon, 
which is in general different from the rest frame 
in the asymptotic region near the spatial infinity.


\subsection{Entanglement entropy in the no-island configuration}\label{ssec:NoIslandStat}

Now, we calculate the entanglement entropy of 
the Hawking radiation in the stationary black hole geometry. 
For simplicity, we first consider the following metric, 
\begin{equation}
 ds^2 = - e^{2\rho} dU\,dV + r^2 \bar\sigma_{ij} d \theta^i d \theta^j \ , 
 \label{MSmetric}
\end{equation}
and discuss further generalization later. 
Here, $\rho$ and $r$ depend only on $U$ and $V$, 
and $\bar \sigma_{ij}$ depend only on the angular coordinates $\theta$. 
Then, the angular dependence can easily be separated, 
and we can focus on the $(U,V)$-dependence in the s-wave approximation. 
Furthermore, the stationary geometry has the Killing vector \eqref{Killing}. 
The components $\rho$ and $r$ depend only on the combination $UV$ and 
can be expanded around the bifurcate horizon as 
\begin{align}
 \rho 
 &= 
 - \rho_1 UV + \mathcal O(U^2V^2) \ , 
\label{rho-exp}
\\
 r 
 &= 
 r_h - r_1 UV + \mathcal O(U^2V^2) \ , 
\label{r-exp}
\end{align}
where we used the condition $\rho(UV=0)=0$, and $r_h$ is the radius on the horizon. 

\begin{figure}[t]
\begin{center}
\raisebox{128pt}{(a)} \includegraphics[scale=0.4]{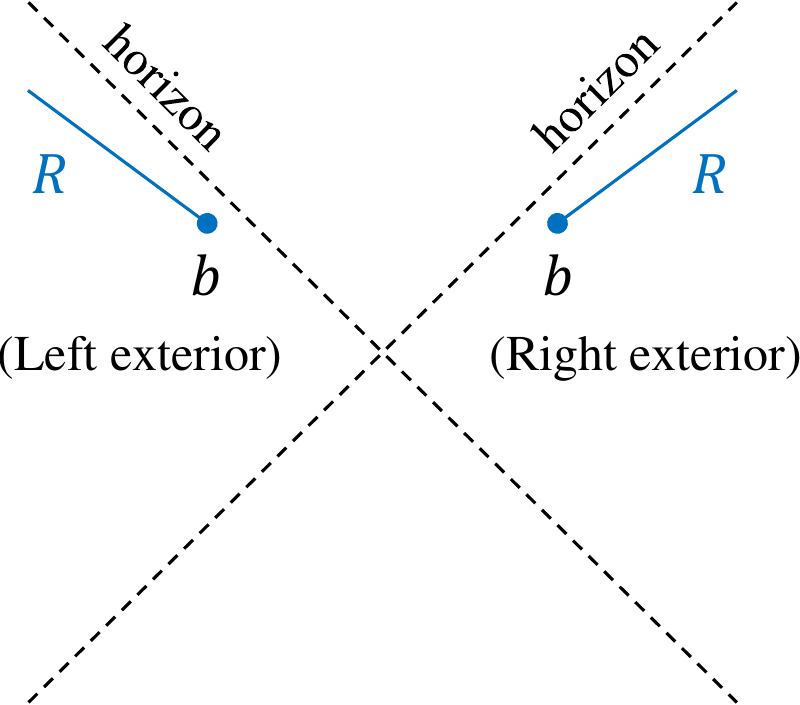}
\hspace{48pt}
\raisebox{128pt}{(b)} \includegraphics[scale=0.4]{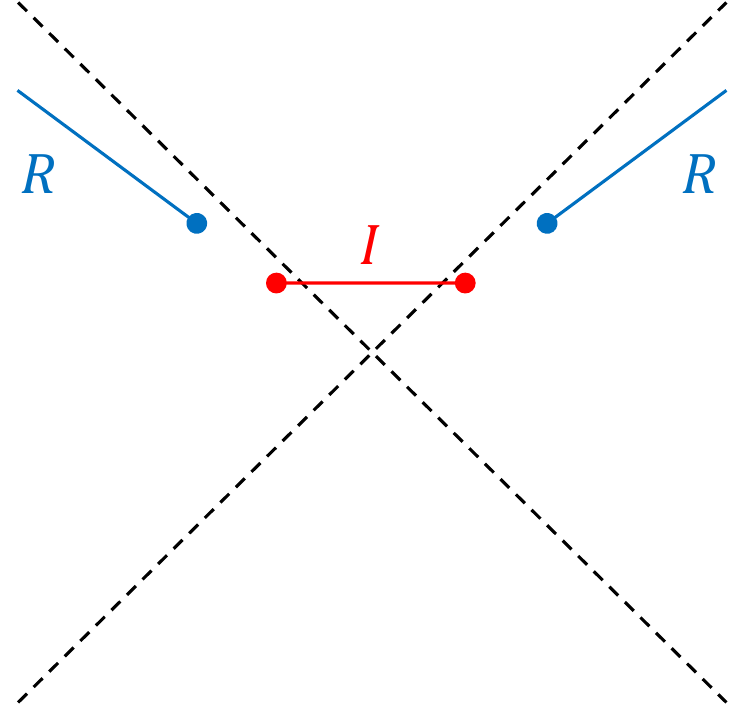}
\caption{%
(a) The bifurcate horizon in stationary black holes (dashed lines) and 
region $R$ in the right and left exteriors. 
(b) The island $I$ extends slightly outside the horizon. 
}\label{fig:stationary}
\end{center}
\end{figure}

As we have discussed in the previous section, 
the entanglement entropy of the Hawking radiation is identified with 
that of a region outside the horizon. 
For stationary black holes, the geometry has two exteriors of the horizon, 
and the Hawking radiation appears in both two exteriors. 
Thus, the region $R$ of the Hawking radiation is union of two regions in two exteriors of the horizon 
(Fig.~\ref{fig:stationary}). 
The exteriors of the horizon is given by $UV<0$, 
and we call $U<0<V$ the right exterior and $V<0<U$ the left exterior.  
The region $R$ is defined as the outside of some boundary $b$ on a timeslice. 

We first calculate the entanglement entropy in the configuration without islands. 
As we discussed in the previous section, we ignore the area term of the boundary of $R$. 
Then, the entanglement entropy is simply given by the matter part. 
In the s-wave approximation, we obtain the two dimensional effective theory 
of matter fields by integrating over $\theta^i$-directions. 
S-waves of massless fields are approximated by two-dimensional massless fields, 
and then, the entanglement entropy is given by using \eqref{Smatter} as%
\footnote{%
Here, we use the $(U,V)$-coordinate, in which the metric near the horizon is given by \eqref{NHmetric1}. 
Then, the entanglement entropy is calculated in 
the Hartle-Hawking vacuum state: $T_{UU}=T_{VV}=0$ near the horizon. 
This condition is necessary for stationary black holes to avoid the divergence on the horizon. 
} 
\begin{equation}
 S 
 = 
 \frac{c}{6} \log\left|\left(U_b-U_b'\right)\left(V_b-V_b'\right)\right| 
 + \frac{c}{3} \rho \ , 
\end{equation}
where $(U_b,V_b)$ stands for the position of $b$ --- the boundary of $R$ in the right exterior. 
We put the boundary of $R$ in the left exterior so that the region $R$ 
is symmetric under the exchange of the left and right exteriors, $U\to V$ and $V\to U$. 
The boundary in the left exterior is given by 
\begin{align}
 U_b' &= V_b \ , 
 & 
 V_b' &= U_b \ . 
 \label{symR}
\end{align}
Thus, the entanglement entropy is expressed as 
\begin{equation}
 S 
 = 
 \frac{c}{3} \log\left|V_b-U_b\right| 
 + \frac{c}{3} \rho \ . 
\end{equation}
We put the boundary of the region $R$ near the horizon $|UV|\ll 1$. 
When the boundary moves along a surface with constant radius, or equivalently $|UV|=\text{const.}$, 
the entanglement entropy behaves as 
\begin{equation}
 S 
 \simeq 
 \frac{c \kappa v_b}{3} \ , 
 \label{NoIstat}
\end{equation}
at sufficiently late time $v_b\gg \kappa^{-1}$, 
where $v_b = \kappa^{-1} \log V_b$ is the advanced time 
associated to the generator of the Killing horizon at $U=0$. 

\subsection{Islands in stationary black holes}\label{ssec:IslandStat}

Next, we consider the configuration with an island. 
The gravity part of the entanglement entropy has four terms: 
two of them are the contributions from $b$ (boundaries of the region $R$) and 
the other two are from boundaries of the island $I$, or equivalently, quantum extremal surfaces $a$. 
Thus, we have to take the gravity part into account though we ignore local terms at $b$ in the gravity part.  
The entanglement entropy is the sum of the gravity part and matter part, 
\begin{equation}
 S = 
 S_\text{grav}(U_a,V_a) 
 + S_\text{grav}(U'_a,V'_a) 
 + S_\text{matter} \ , 
\end{equation}
and the matter part is given by 
\begin{align}
 S_\text{matter} 
 &= 
 \frac{c}{6} \log\left|
   (U_a-U'_a)(V_a-V'_a)
 \right| 
 + \frac{c}{6} \log\left|
   (U_b-U'_b)(V_b-V'_b)
 \right| 
\notag\\
&\quad
 + \frac{c}{6} \log\left|
   (U_a-U_b)(V_a-V_b)
 \right| 
 + \frac{c}{6} \log\left|
   (U'_a-U'_b)(V'_a-V'_b)
 \right| 
\notag\\
&\quad
 - \frac{c}{6} \log\left|
   (U_a-U'_b)(V_a-V'_b)
 \right| 
 - \frac{c}{6} \log\left|
   (U'_a-U_b)(V'_a-V_b)
 \right| 
\notag\\
&\quad
 + \frac{c}{6}\rho(U_a,V_a)
 + \frac{c}{6}\rho(U'_a,V'_a)
 + \frac{c}{6}\rho(U_b,V_b)
 + \frac{c}{6}\rho(U'_b,V'_b) \ , 
\end{align}
where $(U_a,V_a)$ and $(U'_a,V'_a)$ are positions of two boundaries of the island $I$. 
Boundaries of the island is nothing but the quantum extremal surfaces, 
whose position is determined so that 
the entanglement entropy is extremized, 
\begin{align}
 \frac{\partial S}{\partial U_a} &= 0 \ , 
&
 \frac{\partial S}{\partial V_a} &= 0 \ , 
\label{ext}
\\
 \frac{\partial S}{\partial U'_a} &= 0 \ , 
&
 \frac{\partial S}{\partial V'_a} &= 0 \ . 
\end{align}
We solve these equations assuming that the quantum extremal surfaces are located near the horizon 
and see that they have a solution consistent with the assumption. 
The spacetime near the bifurcate horizon is symmetric under the reflection $U\to V$, $V\to U$, 
and we put the boundaries of the region $R$ to respect this symmetry, 
or equivalently to obey the condition  \eqref{symR}, as in the case without islands. 
Then, positions of two quantum extremal surfaces are also symmetric, 
\begin{align}
 U_a' &= V_a \ , 
 & 
 V_a' &= U_a \ , 
 \label{symI}
\end{align}
and the two terms of the gravity part equal each other, 
\begin{equation}
 S_\text{grav}(U_a,V_a) 
 = 
 S_\text{grav}(U'_a,V'_a) \ . 
\end{equation}

For the consistency, the island should be located outside the causal past or future of the region $R$. 
The region $R$ consists of two connected regions: 
one in the right exterior and the other in the left exterior. 
For the region $R$ in the right exterior, we obtain the condition 
\begin{align}
 U_a &> U_b \ , 
 & 
 V_a &< V_b \ . 
\end{align}
For the region $R$ in the left exterior, by using \eqref{symR}, 
we obtain 
\begin{align}
 U_a &< V_b \ , 
 & 
 V_a &> U_b \ . 
\end{align}
We obtain the same conditions for $(U'_a,V'_a)$ with \eqref{symI}. 
Since one of two boundaries of the island satisfies $U_a < V_a$ 
while the other satisfies $U_a > V_a$, 
we can choose $U_a < V_a$ without loss of generality. 
Thus, the position of the quantum extremal surfaces must satisfy 
\begin{equation}
 U_b < U_a < V_a < V_b \ . 
\label{cond-i}
\end{equation}

Since we assume that (coefficients of) the gravitational action is very large, 
the gravity part $S_\text{grav}$ of the entanglement entropy is usually much larger than the matter part, 
and then, the configuration with the island cannot be the dominant saddle point. 
However, the entanglement entropy for the configuration without islands \eqref{NoIstat} 
increases with time and can be larger than $S_\text{grav}$ at late times. 
At such late times, the distance between the right exterior and left exterior 
becomes very large, and the position of $b$ satisfies $V_b \gg V'_b$. 
The condition \eqref{cond-i} can also be written as 
\begin{equation}
 V'_b < V'_a < V_a < V_b \ . 
\end{equation}
At late times, in order to satisfy $V_b \gg V'_b$, we need 
\begin{equation}
 V'_b < V'_a \ll V_a < V_b \ , 
 \label{cond-it}
\end{equation}
or 
\begin{equation}
 V'_b \ll V'_a < V_a \ll V_b \ . 
 \label{cond-if}
\end{equation}
For \eqref{cond-if}, we find 
\begin{align}
 &\log\left|
   (U_a-U_b)(V_a-V_b)
 \right| 
 + \log\left|
   (U'_a-U'_b)(V'_a-V'_b)
 \right| 
\notag\\
&\quad
 - \log\left|
   (U_a-U'_b)(V_a-V'_b)
 \right| 
 - \log\left|
   (U'_a-U_b)(V'_a-V_b)
 \right| 
\notag\\
 &\simeq
 - 2 \log\left|\frac{V_a}{U_a}\right| \ . 
\end{align}
Then, compared with the case without islands, 
the additional terms in the entanglement entropy are given by 
\begin{align}
 &S 
 - \frac{c}{3} \log\left|V_b-U_b\right| 
\notag\\
 &\simeq
 2 S_\text{grav}(U_a,V_a)
 + \frac{c}{3} \log\left|V_a-U_a\right| 
 - \frac{c}{3} \log\left|\frac{V_a}{U_a}\right| \ , 
\end{align}
which can be negative only if $V_a\gg U_a$. 
Thus, the configuration with the island can be the dominant saddle point only for \eqref{cond-it}. 
Then, we find 
\begin{align}
&
 \log\left|
   (U_a-U'_a)(V_a-V'_a)
 \right| 
 + \log\left|
   (U_b-U'_b)(V_b-V'_b)
 \right| 
\notag\\
&\quad
 - \log\left|
   (U_a-U'_b)(V_a-V'_b)
 \right| 
 - \log\left|
   (U'_a-U_b)(V'_a-V_b)
 \right| 
\notag\\
 &= \mathcal O\left(V^{-2}\right) 
\end{align}
is negligible. We will see that the quantum extremal surface satisfies the ansatz $V_a \gg U_a$, later. 

The term $\rho(U_b,V_b)$ is much smaller than $S_\text{grav}$. 
It does not affect the position of the island as it is independent of $(U_a,V_a)$. 
Thus, $\rho(U_b,V_b)$ can also be ignored. 
Then, the entanglement entropy is expressed as 
\begin{align}
 S 
 &= 
 2S_\text{grav}(U_a,V_a) 
 + \frac{c}{3} \log\left|
   (U_a-U_b)(V_a-V_b)
 \right| 
 + \frac{c}{3}\rho(U_a,V_a)
 \ . 
 \label{Istat0}
\end{align}

Although the second and third terms are much smaller than the first term, 
they possibly affect the position of the island. 
The second term, in fact, plays an important role to determine the position, 
but the third term gives only a small correction. 
In the case of the Einstein gravity, for example, 
the gravity part is given by the area term, 
\begin{equation}
 S_\text{grav}(U_a,V_a) 
 = 
 \frac{\mathrm{Area}(U_a,V_a)}{4G_N} 
 = 
 \frac{r^{D-2}(U_a,V_a)\,\Omega_{D-2}}{4 G_N} \ , 
\end{equation}
where $\Omega_{D-2}$ is the volume of the angular directions,%
\footnote{%
If some of the angular directions are non-compact, 
we consider the entanglement entropy par unit volume in non-compact directions. 
} 
\begin{equation}
 \Omega_{D-2} = \int d^{D-2} \theta \, \sqrt{\bar \sigma} \ . 
\end{equation}
Assuming that the boundaries of the island is located near the horizon, 
the area term can be expanded by using \eqref{r-exp} as 
\begin{equation}
 S_\text{grav}(U_a,V_a) 
 = 
 \frac{r_h^{D-2} \Omega_{D-2}}{4 G_N} \left(1 + (D-2)\frac{r_1}{r_h} U_a V_a + \cdots \right) \ . 
\end{equation}
The third term can be expanded similarly by using \eqref{rho-exp}, 
and then, we obtain 
\begin{equation}
 2S_\text{grav} + \frac{c}{3}\rho 
 = \frac{r_h^{D-2} \Omega_{D-2}}{2 G_N} 
 + \left(\frac{(D-2) r_1 r_h^{D-1}\Omega_{D-2}}{2 G_N} + \frac{c}{3}\rho_1\right) U_a V_a + \cdots \ . 
\end{equation}
Provided that the Newton constant is sufficiently smaller than 
the typical length scale of the black hole $r_h$, 
the contribution from $\frac{c}{3}\rho$ is negligible. 

More generally, the gravity part of the entanglement entropy is given by \eqref{S_grav}--\eqref{S_anom}. 
The gravity part can be expressed in terms of the combination $UV$, 
since all metric components depend on $(U,V)$-coordinates only through $UV$. 
Thus, the gravity part $S_\text{grav}$ can be expanded near the horizon as 
\begin{equation}
 S_\text{grav}(U_a,V_a) = A - BU_aV_a + \cdots \ . 
\end{equation}
As we assume, for the validity of the semi-classical treatment, 
that the coefficients of the gravity part is sufficiently large, 
the third term in \eqref{Istat0} can be ignored. 
 
Now, the entanglement entropy is expressed as 
\begin{equation}
 S = 2S_\text{grav}(U_a,V_a) + \frac{c}{3} \log\left|(U_a-U_b)(V_a-V_b)\right| \ . 
\end{equation}
The conditions of the quantum extremal surface \eqref{ext} become 
\begin{align}
 0 
 &= 
 \frac{\partial S}{\partial U_a} 
 = 
 - 2 B V_a + \frac{c}{3}\frac{1}{U_a-U_b} \ , 
\label{extUs}
\\
 0 
 &= 
 \frac{\partial S}{\partial V_a} 
 = 
 - 2 B U_a + \frac{c}{3}\frac{1}{V_a-V_b} \ . 
\label{extVs}
\end{align}
The condition \eqref{extUs} and \eqref{extVs} for the quantum extremal surface 
have solutions which satisfy \eqref{cond-i} only if $B>0$. 
Therefore, we assume 
\begin{equation}
 B > 0 \ . 
\end{equation}
This condition actually satisfied by the area term 
as the area is larger in the exterior of the horizon $UV<0$. 
For models with $B<0$, no island would appear even at late times, 
and hence, we do not consider such cases. 

The conditions \eqref{extUs} and \eqref{extVs} are solved as 
\begin{align}
 U_a 
 &= 
 \frac{1}{2B V_b}
 \left(
  BU_b V_b \pm \sqrt{B^2 U_b^2 V_b^2 + \tfrac{2}{3} c B U_b V_b}
 \right) \ , 
\\
 V_a 
 &= 
 \frac{1}{2B U_b}
 \left(
  BU_b V_b \pm \sqrt{B^2 U_b^2 V_b^2 + \tfrac{2}{3} c B U_b V_b}
 \right) \ . 
\end{align}
For $|BU_bV_b|\gg c$, these two solutions are approximated as 
\begin{align}
 U_a 
 &\simeq
 - \frac{c}{6 B V_b} \ , 
&
 V_a 
 &\simeq 
 - \frac{c}{6 B U_b} \ , 
\label{sol-et} 
\end{align}
and 
\begin{align}
 U_a 
 &\simeq 
 U_b + \frac{c}{6 B V_b} \ , 
&
 V_a 
 &\simeq 
 V_b + \frac{c}{6 B U_b} \ . 
\label{fake-et}
\end{align}
Both of these two solutions actually satisfy the ansatz $V_a \gg U_a$. 
The solution \eqref{sol-et} gives an extremum of the entanglement entropy, 
\begin{equation}
 S = 2A + \mathcal O(B^{0}) \ ,  
\label{entropy-et} 
\end{equation}
which is nothing but (twice of) the black hole entropy, 
and the other solution \eqref{fake-et} gives 
\begin{equation}
 S = 2A - 2B U_b V_b 
 + \mathcal O(B^0) \ , 
\end{equation}
which is the gravitational entropy at the boundaries of the region $R$. 
The dominant saddle point gives the minimum in the extrema of the entanglement entropy,  
and hence, the solution \eqref{sol-et} gives the position of the quantum extremal surface. 
Thus, the entanglement entropy of the Hawking radiation is given by \eqref{entropy-et} at late times, 
and hence, is bounded from above by (twice of) the black hole entropy $S_\text{grav} \simeq S_\text{Wald}$. 
In the case of the Einstein gravity, $B = \mathcal O(G_N^{-1})$, and hence, 
the quantum extremal surface is located within a Planck distance from the horizon.%
\footnote{%
The square of the distance from the horizon is estimated as $|U_aV_a| = \mathcal O(G_N^{-2})$. 
Note that the distance is not of the same order to the Planck length but much smaller. 
} 
Since $U_b V_b < 0$, the quantum etremal surface is also located in 
$U_a V_a < 0$, namely, outside the horizon. 
It should be noted that the size of the island $I$, or equivalently, 
the distance between two boundaries of $I$ is very large as $V_a\gg U_a$.


\subsection{Further generalization}\label{ssec:General}


\subsubsection{Beyond the s-wave approximation}\label{sssec:GeneralMatter}

So far, we have calculated the entanglement entropy by using the s-wave approximation. 
In the configuration without islands, the approximation is always good 
because the region $R$ in the right exterior is sufficiently separated 
from the region $R$ in the left exterior. 
Correlations between two regions are suppressed 
for modes with (angular) momenta in $\theta$-directions, 
and only massless modes contribute to the entanglement entropy. 
In the configuration with the island, 
the s-wave approximation is good only if 
the region $R$ is sufficiently away from the horizon. 
If (the boundary of) the region $R$ is very close to the horizon, 
we cannot ignore modes with (angular) momenta and 
have to take them into the calculation. 
For massless fields in $D$-dimensional spacetimes, 
the matter part of the entanglement entropy behaves as \cite{Casini:2005zv,Casini:2009sr}
\begin{equation}
 S_\text{matter} 
 \propto 
 - \frac{\text{Area}}{L^{D-2}} \ , 
\label{Smatter-4D}
\end{equation}
where $L$ is the distance between the region $R$ and the island. 

The matter part of the entanglement entropy $S_\text{matter}$ mainly depends on 
the distance between the region $R$ and the island. 
The position of the horizon has almost nothing to do with $S_\text{matter}$
since the spacetime near the horizon can be treated as an approximately flat spacetime. 
An important property of $S_\text{matter}$ is 
\begin{equation}
 \frac{\partial}{\partial L} S_\text{matter} > 0 \ , 
\label{cond-Sm}
\end{equation}
since the interval between the region $R$ and island $I$ is sufficiently separated from the other intervals 
even if the boundaries of the island is very near the horizon, 
as we have seen in the case of the s-wave approximation.  

For simplicity, we utilize the stationary nature of the spacetime. 
We introduce the $(t,x)$-coordinates as 
\begin{align}
 U 
 &=
 - e^{-\kappa t} x \ , 
 &
 V 
 &= 
 e^{\kappa t} x \ , 
\end{align}
where $t$ is the time coordinate associated to the timelike Killing vector, 
and $x$ is the proper coordinate in the radial direction. 
The metric \eqref{MSmetric} is now expressed as 
\begin{equation}
 ds^2 
 = 
 e^{2\rho} \left(-\kappa^2 x^2 dt^2 + dx^2\right) 
 + r^2 \bar\sigma_{ij} d \theta^i d \theta^j \ . 
\end{equation}
The horizon is located at $x=0$, and this set of coordinates 
covers only both exteriors of the horizon. 
Around the horizon, $\rho$ and $r$ are expanded as 
\begin{align}
 \rho 
 &= 
 \rho_1 x^2 + \mathcal O(x^4) \ , 
\\
 r 
 &= 
 r_h + r_1 x^2 + \mathcal O(x^4) \ . 
\end{align}

We assume that the quantum extremal surface is located near the horizon, 
and then, the matter part of the entanglement entropy is expanded as 
\begin{equation}
 S_\text{matter}(x_a,x_b) 
 = S_0 - D_m x_a + \mathcal O(x_a^2) \ , 
\end{equation}
where we focused only on the $x_a$-dependence%
\footnote{%
Although we do not write explicitly, $S_0$ and $D_m$ depend on $x_b$. 
}
of $S_\text{matter}$ and defined 
\begin{align}
 S_0 &= \left.S_\text{matter}\right|_{x_a=0}
 \\
 D_m &= - \left.\frac{\partial}{\partial x_a}S_\text{matter} \right|_{x_a=0} \ . 
\end{align}
Then, the condition \eqref{cond-Sm} implies 
\begin{equation}
 D_m > 0 \ . 
\end{equation}
The gravity part of the entanglement entropy $S_\text{grav}$ is also expanded as 
\begin{equation}
 S_\text{grav}(x_a) = A + B x_a^2 + \cdots \ . 
\end{equation}
Thus, the position of the quantum extremal surface is determined by 
\begin{equation}
 0 
 = 
 \frac{\partial S}{\partial x_a} 
 = 
 4 B x_a - D_m \ ,  
\end{equation}
and is given by 
\begin{equation}
 x_a = \frac{D_m}{4B} \ , 
 \label{sol-GenMat}
\end{equation}
which is actually very close to the horizon. 
In the case of the Einstein gravity, we have 
$B = \mathcal O(G_N^{-1})$ while $D_m = \mathcal O(G_N^0)$. 
Thus, the distance from the horizon is smaller than the Planck length. 
By substituting \eqref{sol-GenMat}, 
the entanglement entropy is obtained as 
\begin{equation}
 S = 2A + S_0 + \mathcal O(B^{-1}) = 2A + \mathcal O(B^0) \ . 
\end{equation}
Thus, the entanglement entropy is approximately given by 
(twice of) the Bekenstein-Hawking entropy. 

The position \eqref{sol-GenMat} of the quantum extremal surface is 
given in the coordinate $x$, which covers only outside the horizon, 
implying that the quantum extremal surface is located outside the horizon. 
In order to see that there is no other saddle point inside the horizon, 
we need to calculate the extremality condition \eqref{ext} 
in coordinates which cover the interior of the horizon. 
In $(U,V)$-coordinates, the condition \eqref{cond-Sm} implies 
\begin{align}
 \frac{\partial}{\partial U_a} S_\text{matter} &> 0 \ , 
&
 \frac{\partial}{\partial V_a} S_\text{matter} &< 0 \ .  
\label{cond-Sm-uv}
\end{align}
Then, the condition \eqref{ext} can be written as 
\begin{align}
 0 
 &= 
 \frac{\partial S}{\partial U_a} 
 = 
 - 2 B V_a + \frac{\partial S_\text{matter}}{\partial U_a} \ , 
\\
 0 
 &= 
 \frac{\partial S}{\partial V_a} 
 = 
 - 2 B U_a + \frac{\partial S_\text{matter}}{\partial V_a} \ , 
\end{align}
if the quantum extremal surface is near the horizon. 
For \eqref{cond-Sm-uv}, these equations can be satisfied only 
for $U_a<0$ and $V_a>0$, namely, outside the horizon. 
If $(U_a,V_a)$ is not near the horizon, 
terms of the gravity part in \eqref{ext} cannot be comparable with the matter part, 
assuming that there is no classical extremal surface other than the bifurcation surface. 
Therefore, there is no saddle point of the quantum extremal surface inside the horizon. 

The calculation above implies that the quantum extremal surface 
is obtained by a small shift of the classical extremal surface 
which is located at the bifurcation surface of the horizon $x=0$. 
As the matter part of the entanglement entropy is much smaller than 
the gravity part, the shift of the extremal surface in each side is also very small. 
However, two quantum extremal surfaces are located in the right and left exteriors respectively, 
and the distance between them can be very large 
though the distance from the bifurcation surface is very small. 
Thus, the island has a very large volume 
although it arises from a small quantum correction by the matter part. 

So far, we assumed that the matter part of the entanglement entropy 
$S_\text{matter}$ is of the same order to the number of the degrees of freedom. 
If the size of the interval, the distance between the region $R$ and island in our case, 
approaches zero, or the cut-off scale to be more precise, 
the total entanglement entropy approaches to zero. 
The matter part originally contains divergent terms, 
which are absorbed by the gravity part as the renormalization of gravitational coupling constants. 
Thus, the divergent terms are effectively subtracted 
from our matter part $S_\text{matter}$, 
and hence, $S_\text{matter}$ becomes very large negative 
when the size of the interval approaches the cut-off scale. 
In this case, the distance between the quantum extremal surface and the horizon will be larger, 
as the quantum extremal surface is moved from 
the bifurcation surface of the horizon due to the contribution from $S_\text{matter}$. 

For example, in the case of a $D$-dimensional spacetime, 
the matter part \eqref{Smatter-4D} becomes 
\begin{equation}
 S_\text{matter} \sim \frac{1}{(x_b - x_a)^{D-2}} \sim \frac{1}{x_a^{D-2}} \ , 
\end{equation}
when the interval $L = x_b - x_a$ is of the same order to the distance between 
the quantum extremal surface and horizon, namely $x_b \sim x_a$.%
\footnote{%
The saddle point with $L\ll x_a$ does not give the minimum of the entanglement entropy. 
For example, by using the s-wave approximation in the Einstein gravity, 
we obtain $x_a = \mathcal O(G_N^0)$ for $L \sim G_N$. 
In this case, the distance between the region $R$ and horizon 
is estimated as $x_b \sim L + x_a = \mathcal O(G_N^0)$, 
for which we have another saddle point with $x_a = \mathcal O(G_N)$ and $L =\mathcal O (G_N^0)$. 
These two saddle points correspond to \eqref{fake-et} and \eqref{sol-et}, respectively. 
} 
The derivative of $S_\text{matter}$ behaves as 
\begin{equation}
 \frac{\partial}{\partial L} S_\text{matter} 
 \sim 
 \frac{1}{(x_b - x_a)^{D-1}} 
 \sim 
 \frac{1}{x_a^{D-1}} \ , 
\end{equation}
and then, the proper distance between the quantum extremal surface and the horizon is 
roughly estimated as 
\begin{equation}
 x_a^{D} \sim G_N \ , 
\end{equation}
in the case of the Einstein gravity. 
Thus, the distance between the quantum extremal surface and horizon 
can be larger than the Planck length if the region $R$ is very close to the horizon 
\cite{Matsuo:2020ypv,Bousso:2023kdj}.


\subsubsection{More general black holes}\label{sssec:GeneralBH}

So far, we have studied black holes whose metric take the form of \eqref{MSmetric}. 
A simplest generalization of \eqref{MSmetric} would be to introduce 
the rotation in the following form, 
\begin{equation}
 ds^2 = - e^{2\rho(x)} dU\,dV 
 + r^2(x) \bar\sigma_{ij}(\theta) 
 \left(d \theta^i + A^i_\alpha (x) dx^\alpha\right) 
 \left(d \theta^j + A^j_\beta (x) dx^\beta\right) 
 \ . 
\end{equation}
where $x^\alpha=(U,V)$ is the temporal and radial coordinates, 
and $\theta$ is the angular coordinates. 
The off-diagonal components of the metric $A^i_\alpha$ simply 
behave as the Kaluza-Klein gauge fields in the two-dimensional effective theory 
for s-waves of matter fields. 
By using an appropriate angular coordinates, 
the Kaluza-Klein gauge fields $A^i_\alpha$ obey the condition \eqref{NHgauge} at the horizon, 
and then, the metric near the horizon is given by \eqref{NHmetric1}. 

The gravity part of the entanglement entropy is the same to the case without rotation. 
After the dimensional reduction to the two-dimensional effective theory, 
the Kaluza-Klein gauge fields $A^i_\alpha$ are treated as matter fields 
and contribute to the matter part of the entanglement entropy. 
As long as they have no singular behavior at the horizon, 
the general analysis in Sec.~\ref{sssec:GeneralMatter} can be applied 
also for the Kaluza-Klein gauge fields and matter fields with the Kaluza-Klein charges. 
Thus, the quantum extremal surface is located very close to but outside the horizon, 
as we have discussed in Sec.~\ref{sssec:GeneralMatter}. 

For more general black holes, the angular dependence and radial dependence may not be factorized. 
The metric is expressed as \eqref{STmetric}, and 
the components may depend on the radial and angular coordinates. 
By using the time coordinate associated to the Killing vector of 
the stationary geometry, the metric is expressed as 
\begin{equation}
 ds^2 
 = 
 e^{2\rho(x,\theta)} \left(- \kappa^2 x^2 dt^2 + dx^2\right)
 + \sigma_{ij}(x,\theta) 
 \left(d \theta^i + A^i_\alpha(x,\theta) dx^\alpha\right) 
 \left(d \theta^j + A^j_\beta(x,\theta) dx^\beta\right) \ , 
\end{equation}
where $x^\alpha = (t,x)$. 
In such cases, the radial position of the quantum extremal surface, 
$x=x_a(\theta)$, depends on the angular position. 
Here, we focus on the Einstein gravity for simplicity, 
but the generalization to general gravity theories is straightforward. 
The gravity part of the entanglement entropy is expressed as 
\begin{equation}
 S_\text{grav} 
 = 
 \frac{1}{4G_N}\int d^{D-2}\theta\sqrt{\Sigma} \ , 
\end{equation}
where $\Sigma_{ij}$ is the induced metric on the quantum extremal surface, 
which is expressed in $\theta$-coordinates as 
\begin{equation}
 \Sigma_{ij} 
 = 
 e^{2\rho} \left(\frac{\partial x}{\partial \theta^i}\right)
 \left(\frac{\partial x}{\partial \theta^j}\right) 
 + \sigma_{kl} 
 \left[\delta^k{}_i + A^k_x \left(\frac{\partial x}{\partial \theta^i}\right)\right] 
 \left[\delta^l{}_j + A^i_x \left(\frac{\partial x}{\partial \theta^j}\right)\right] 
\ . 
\end{equation}

The matter part of the entanglement entropy depends 
on the position of $a$, the quantum extremal surface, and $b$, the boundary of region $R$. 
In the case of the symmetric space \eqref{MSmetric}, 
the positions of these surfaces can be identified by points 
in the two-dimensional spacetime of temporal and radial directions. 
The summation in the formula \eqref{Smatter} would be replaced by 
the integration over the surface as 
\begin{equation}
 S_\text{matter}(a,b) 
 = 
 \int_a d^{D-2}\theta \sqrt{\Sigma}\, s_m(x_a(\theta), \theta) \ , 
\end{equation}
where we focused only on the dependence on the position of 
the quantum extremal surface $a$, and the integration is on the surface $a$. 
The integrand $s_m$ contains integration over the position of $b$. 
We do not specify the concrete form of $s_m$ but assume 
\begin{equation}
 \frac{\partial s_m}{\partial x_a} < 0 \ , 
\end{equation}
similarly to the case in Sec.~\ref{sssec:GeneralMatter}. 

The metric is expanded near the horizon, or equivalently for small $x$ as 
\begin{align}
 \rho 
 &= 
 \rho_1(\theta) x^2 + \cdots \ , 
\\
 \sigma_{ij} 
 &= 
 r_h^2 \bar \sigma_{ij}(\theta) 
 + \sigma^{(1)}_{ij}(\theta) x^2 + \cdots \ . 
\end{align}
The integrand of the matter part is also expanded as 
\begin{equation}
 s_m(x,\theta) = s_0(\theta) + d_m(\theta) x + \mathcal O(x^2) \ .  
\end{equation}
Thus, the condition for the quantum extremal surface is given by 
the following differential equation for $x_a$, 
\begin{equation}
 0 
 = 
 \frac{\delta S}{\delta x_a(\theta)} 
 = 
 \frac{1}{2G_N} \sigma^{(1)}(\theta) x_a(\theta) - \frac{1}{2G_N} \nabla^2 x_a(\theta) - d_m(\theta) \ , 
\label{ext-gen}
\end{equation}
where $\nabla_i$ is the covariant derivative on (the time slice of) the horizon, 
and $\sigma^{(1)} = (\sigma^{(1)})^i{}_i$. 
The differential equation above cannot be solved for 
an arbitrary near horizon metric $\sigma_{ij}$ and arbitrary matter $d_m(\theta)$. 
We make a rough estimation of the solution by using the following ansatz, 
\begin{equation}
 x_a(\theta) = x_a^{(0)}(\theta) + \Delta x_a(\theta) \ , 
\end{equation}
where $x_a^{(0)}$ is given by 
\begin{equation}
 x_a^{(0)}(\theta) = \frac{2G_N d_m(\theta)}{\sigma^{(1)}(\theta)} \ . 
\end{equation}

If $\Delta x_a$ is negligible, the quantum extremal surface is very close to the horizon 
as $x_a = \mathcal O(G_N)$ and is located outside the horizon. 
The entanglement entropy is then approximately given by 
\begin{equation}
 S = 2A + \mathcal O(G_N^0) \ . 
\end{equation}
The correction to this naive approximation is given by 
\begin{equation}
 \Delta x_a(\theta) 
 = 
 \frac{\nabla^2 x_a(\theta)}{\sigma^{(1)}(\theta)} \ . 
\end{equation}
Provided that the solution of \eqref{ext-gen} is smooth, 
the second term obeys at the maximum of $x_a$ 
\begin{equation}
 \nabla^2 x_a < 0 \ , 
\end{equation}
and hence, the correction term is 
\begin{equation}
 \Delta x_a < 0 \ . 
\end{equation}
Thus, maximum of the solution is smaller than $x_a^{(0)}$. 
Similarly, at the minimum of $x_a$, the second term satisfies 
\begin{equation}
 \nabla^2 x_a > 0 \ , 
\end{equation}
and hence, the correction term is 
\begin{equation}
 \Delta x_a > 0 \ . 
\end{equation}
Thus, the solution of \eqref{ext-gen} satisfies 
\begin{equation}
 \min x_a^{(0)} < x_a < \max x_a^{(0)} \ , 
\end{equation}
and hence, $x_a = \mathcal O(G_N)$ and $x_a>0$. 
Therefore, the quantum extremal surface is located very close to but outside the horizon, 
and the entanglement entropy is approximately given by the black hole entropy, 
\begin{equation}
 S = 2A + \mathcal O(G_N^0) \ . 
\end{equation}


\section{Islands in evaporating black holes}\label{sec:Evaporating}

In the previous section, we studied the entanglement entropy 
of the Hawking radiation in stationary black holes, 
and found that the island extends slightly outside the horizon. 
In this section, we consider evaporating black holes 
to see that the island is hidden by the horizon, 
or equivalently, the quantum extremal surface is inside the horizon. 


\subsection{Evaporating black holes and Hawking radiation}\label{ssec:Hawking}

In this section, we study black holes which are evaporating by emitting the Hawking radiation. 
In contrast to the stationary black holes, 
evaporating black boles are dynamical, and the metric has time dependence. 
The metric in the vicinity of the horizon at $U=0$ is given by \eqref{NHmetric}, 
but the surface gravity $\kappa$ and the horizon radius depend on the advanced time $v$.%
\footnote{%
To be precise, the metric near the apparent horizon would be approximated by 
the metric of stationary black holes for sufficiently shorter time period 
than the lifetime of the black hole. 
The event horizon is a null surface which can be chosen to be $U=0$, 
while the apparent horizon is a timelike surface for evaporating black holes. 
The event horizon would be located sufficiently near the apparent horizon, 
and then, the metric near the event horizon can also be approximated by the stationary metric. 
} 

The surface gravity appears in the geodesic equation \eqref{geodesic} 
as the advanced time $v$ is not an affine parameter. 
The relation between the advanced time $v$ and an affine parameter $V$ 
is expressed as 
\begin{equation}
 dV = \exp\left[\int^v dv' \kappa(v') \right] dv \ , 
 \label{V-eva}
\end{equation}
and hence, $C(U,v)$ in the metric \eqref{NHmetric} is given by 
\begin{equation}
 C(U,v) \simeq \exp\left[\int_{v_0}^v dv' \kappa(v') \right] \ , 
 \label{c-eva}
\end{equation}
up to an overall factor. 
Here, $v_0$ is a reference time, which we take around the time of the black hole formation 
so that the metric \eqref{NHmetric} can be connected 
to the static spacetime with $\kappa=0$, 
\begin{equation}
 ds^2 = - dU\,dv + \sigma_{ij} d \theta^i d \theta^j \ , 
\end{equation}
around $v=v_0$, since an empty spacetime would have $\kappa=0$. 
We consider a time scale comparable with the lifetime of the black hole or the Page time, 
and then, the overall factor of \eqref{c-eva} gives 
only subleading corrections to the integral in \eqref{c-eva}, namely 
\begin{equation}
 \int_{v_0}^v dv' \kappa(v') = \mathcal O(G_N^{-1}) \ , 
\end{equation}
in the case of the Einstein gravity for example, 
and then, 
\begin{equation}
 C(U,v) \simeq \exp\left[\int_{v_0}^v dv' \kappa(v') + \mathcal O(G_N^0)\right] \ . 
\end{equation}
In a similar fashion to $\kappa(v)$, 
the horizon radius $r_h$ of evaporating black holes also has $v$-dependence. 
The horizon radius $r_h$ appears in the induced metric on a time slice of the horizon as 
\begin{equation}
 \sigma_{ij}(U=0,v,\theta) = r_h^2(v) \, \bar \sigma_{ij}(\theta) \ . 
\end{equation}
Here, we assume that the form of the horizon is independent of time
but only its size changes with time. 
Thus, the metric very near the horizon of an evaporating black hole is approximately given by  
\begin{equation}
 ds^2 
 \simeq 
 - \exp\left[\int_{v_0}^v dv' \kappa(v')\right] dU\,dv 
 + r_h^2(v) \bar \sigma_{ij}(\theta) d \theta^i d \theta^j \ . 
\end{equation}

Similarly to the case of stationary black holes, 
the metric near the horizon can be expressed in the form of the expansion around $U=0$. 
The time evolution of evaporating black holes is relevant 
only in time scales comparable to the lifetime of the black hole. 
The metric is approximately the same to the stationary black holes 
\eqref{MSmetric}--\eqref{r-exp} for smaller time periods, 
but coefficients in \eqref{rho-exp} and \eqref{r-exp} have slow $v$-dependence. 
By using \eqref{V-eva}, or equivalently \eqref{NHmetric} with \eqref{c-eva}, 
the metric in the $(U,v)$-coordinates is expressed as%
\footnote{%
The definition of $\rho$ here is different from one in the previous section, namely in \eqref{MSmetric}. 
In the previous section, $\rho$ is defined as the conformal factor in the $(U,V)$-coordinates, 
but here, $\rho$ is chosen to be the conformal factor in the $(U,v)$-coordinates, 
so that it agrees with $\rho$ in \eqref{Smatter} in the appropriate vacuum state in each case. 
} 
\begin{equation}
 ds^2 = - e^{2\rho} dU\,dv + r^2 \bar \sigma_{ij} d \theta^i d \theta^j \ , 
\end{equation}
where 
\begin{align}
 \rho 
 &= 
 \frac{1}{2}\int_{v_0}^v dv' \kappa(v') 
 + \rho_1(v) U \exp\left[\int_{v_0}^v dv' \kappa(v')\right] + \cdots \ , 
\label{rho-dyn}
\\
 r 
 &= 
 r_h(v) + r_1(v) U \exp\left[\int_{v_0}^v dv' \kappa(v')\right] + \cdots \ , 
\label{r-dyn}
\end{align}
near the horizon. 
The $v$-dependence of $\kappa$, $r_h$, $\rho_1$ and $r_1$ are very slow, namely, 
\begin{align}
 \dot \kappa &\equiv \frac{d \kappa}{dv} \ll \kappa^2 \ , 
\\
 \dot r_h &\equiv \frac{dr_h}{dv} \ll 1 \ , 
\end{align}
and similarly for $\rho_1$ and $r_1$. 

Next, we calculate the energy-momentum tensor in the black hole geometry. 
We apply the s-wave approximation, in which 
matter fields are approximated by two-dimensional fields. 
The two-dimensional energy-momentum tensor is obtained 
by integrating over all angular coordinates $\theta^i$, 
\begin{equation}
 T^\text{(2D)}_{\mu\nu} = \int d^{D-2}\theta \sqrt{\sigma}\,T_{\mu\nu} \ . 
\end{equation}
The energy-momentum tensor of two-dimensional massless fields 
can be evaluated by using the the conservation law, 
\begin{equation}
 \nabla_\mu T^{\text{(2D)}\,\mu\nu} = 0 \ , 
\end{equation}
and the condition of the trace anomaly, 
\begin{equation}
 T^{\text{(2D)}\,\mu}{}_\mu = \frac{c}{24\pi} R^\text{(2D)} \ , 
\end{equation}
where $R^\text{(2D)}$ is the scalar curvature of 
the two-dimensional spacetime after integrating out the angular directions. 
The energy-momentum tensor is calculated as \cite{Davies:1976ei,Davies:1976hi} 
\begin{align}
 T^\text{(2D)}_{UU} 
 &=
 - \frac{c}{12\pi} C^{1/2} \partial_U^2 C^{-1/2} + \frac{c}{24\pi} F(U) \ , 
\label{Tuu-vac}
\\
 T^\text{(2D)}_{vv} 
 &=
 - \frac{c}{12\pi} C^{1/2} \partial_v^2 C^{-1/2} + \frac{c}{24\pi} \bar F(v) \ , 
\label{Tvv-vac}
\\
 T^\text{(2D)}_{Uv} 
 &=
 - \frac{c}{24\pi}
 \left[
  C^{-1} \partial_U\partial_v C 
  - C^{-2} \left( \partial_U C \right) \left( \partial_v C \right)
 \right] \ ,
\label{Tuv-vac}
\end{align}
where $F(U)$ and $\bar F(v)$ are integration constants. 
Although the energy-momentum tensor must be covariant under the coordinate transformation, 
the expressions above is not written in covariant forms, 
implying that the integration constants transform non-trivially. 
Except for the integration constants, the energy-momentum tensor above is 
completely determined by the anomaly condition and conservation law, 
implying that it depends on the physical state only through the integration constants. 

In the case of evaporating black holes, 
the integration constants above are determined by the initial condition. 
We assume that there is no incoming energy after the formation of the black hole 
and that there is no outgoing energy just after the formation. 
The state with these initial conditions is approximately given by the Unruh vacuum, 
and then, the integration constants are 
\begin{equation}
 F(U) = \bar F(v) = 0 \ , 
\end{equation}
in the $(U,v)$-coordinates in \eqref{NHmetric}. 
Substituting the explicit form of the near horizon metric \eqref{NHmetric} 
into the expressions \eqref{Tuu-vac}--\eqref{Tuv-vac}, 
the energy-momentum tensor near the horizon is estimated as 
\begin{align}
 T^\text{(2D)}_{vv} 
 &\simeq  
 - \frac{c \kappa^2}{48\pi} \ , 
\label{Tvv}
\\
 T^\text{(2D)}_{UU} 
 &= 
 T^\text{(2D)}_{Uv} 
 = 0 \ . 
\end{align}
These expressions imply that there is no outgoing energy near the horizon, 
but the black hole loses the energy due to the incoming negative energy. 
Then, the loss of the energy equals to the incoming negative energy \eqref{Tvv} and is given by 
\begin{equation}
 \frac{dE}{dv} = - \frac{c \kappa^2}{48\pi} \ . 
\end{equation}

The first law of thermodynamics gives the relation between 
the black hole entropy $\S_\text{BH}$ and the energy $E$ as 
\begin{equation}
 \frac{\kappa}{2\pi} \delta S_\text{BH} = \delta E \ . 
 \label{thermo}
\end{equation}
Then, the black hole entropy decreases with time as 
\begin{equation}
 \frac{d S_\text{BH}}{dv} = - \frac{c \kappa}{24} \ . 
\label{dSBH}
\end{equation}

In the case of the Einstein gravity, $S_\text{BH}$ is given by the area of the horizon, 
\begin{equation}
 S_\text{BH} = \frac{r_h^{D-2} \Omega_{D-2}}{4G_N} \ , 
\end{equation}
and then, the derivative of the horizon radius is given by 
\begin{equation}
 \frac{dr_h}{dv} = - \frac{c \, G_N \kappa}{6 (D-2) r_h^{D-3} \Omega_{D-2}} \ . 
\end{equation}


\subsection{Entanglement entropy in the no-island configuration}\label{ssec:NoIslandDyn}

We calculate the entanglement entropy of the Hawking radiation around an evaporating black holes 
in the configuration without islands. 
As we ignore the gravity part which comes from the boundary of the region $R$, 
the entanglement entropy is simply given by the matter part \eqref{Smatter} 
in the s-wave approximation. 
The expression \eqref{Smatter} depends on coordinates, 
which should be chosen to be those associated to the vacuum state (See Appendix). 
Here, we consider the Unruh vacuum, and hence, 
use \eqref{Smatter} with $(U,v)$-coordinates. 
Thus, the entanglement entropy in the configuration without islands is expressed as  
\begin{equation}
 S = \frac{c}{6} 
 \log\left|
  \left( U_b - U'_b \right) \left( v_b - v'_b \right)
 \right|
 + \frac{c}{3}\rho(U_b,v_b) \ , 
 \label{NoIdyn0}
\end{equation}
where $\rho$ is given by \eqref{rho-dyn}, and 
$(U_b,v_b)$ is the position of the boundary of the region $R$.  

If the geometry of evaporating black holes are extended to the infinite past $v\to-\infty$, 
the mass of the black hole will be infinity in $v\to-\infty$. 
Hence, we consider either of the following two cases: 
the black hole is formed around $v=v_0$ or stationary until $v=v_0$. 
If the black hole is formed around $v=v_0$, 
it is so-called the one-sided spacetime, or equivalently, has no Einstein-Rosen bridge. 
In this case, the geometry has the origin, at which incoming modes turn into outgoing modes. 
We introduce the mirror image at the origin, 
and $(U'_b,v'_b)$ is the position of the boundary of the region $R$ in the mirror image 
in terms of the coordinates in the original spacetime, which is expressed as 
\begin{align}
 U'_b 
 &\simeq 
 v_b - v_m \ , 
\\
 v'_b 
 &\simeq 
 U_b + v_m \ , 
\end{align}
where $v_m$ is the advanced time of the intersection of the horizon and origin, 
and $v_0-v_m$ is comparable to the size of the black hole. 
We consider the time evolution of the entanglement entropy 
with the boundary of the region $R$ kept near the horizon, 
and then, $U_b$ becomes very small at sufficiently late times. 
We find 
\begin{equation}
 \frac{c}{6} 
 \log\left|
  \left( U_b - U'_b \right) \left( v_b - v'_b \right)
 \right| 
 \simeq 
 \frac{c}{3} 
 \log\left|
  v_b - v_0 
 \right| \ . 
\end{equation}
Also, the second term of \eqref{NoIdyn0} is calculated by using \eqref{rho-dyn} 
and approximately given by 
\begin{equation}
 \frac{c}{3}\rho(U_b,v_b) \simeq \frac{c}{6}\int_{v_0}^{v_b} dv\, \kappa(v) \ . 
\end{equation}
In the case of one-sided spacetimes, \eqref{NoIdyn0} is 
the entanglement entropy of Hawking radiation and its mirror image, 
which is twice of the true entanglement entropy. 
Thus, the entanglement entropy is calculated as 
\begin{equation}
 S \simeq 
 \frac{c}{6} 
 \log\left|
  v_b - v_0 
 \right| 
 + \frac{c}{12}\int_{v_0}^{v_b} dv\, \kappa(v) 
 \simeq \frac{c}{12}\int_{v_0}^{v_b} dv\, \kappa(v) \ . 
\label{NoIeva}
\end{equation}

Instead of the gravitational collapse, 
we can consider a stationary black hole 
which starts to evaporates at some time $v=v_0$. 
The black hole emits the Hawking radiation when quantum effects are taken into account, 
and hence, we have to introduce incoming radiation for stationary black holes. 
Thus, the vacuum state would be the Hartle-Hawking vacuum for $v<v_0$. 
The incoming radiation is turned off at $v=v_0$, and 
the vacuum state becomes the Unruh vacuum for $v>v_0$.%
\footnote{%
Since the past horizon is located in $v\to -\infty$, 
the regularity condition there is satisfied by taking the Hartle-Hawking vacuum for $v<v_0$. 
} 
Then, the entanglement entropy \eqref{Smatter} should be written in terms of 
the coordinates $(U,v)$ for $v>v_0$ and $(U,V)$ for $v<v_0$ up to a constant shift, where 
\begin{equation}
 V = V_0 \exp\left[\int_{v_0}^v dv' \kappa(v')\right] 
\end{equation}
is defined so that $V=0$ at the past horizon. 

If the black hole is stationary until $v=v_0$, 
the spacetime has two exteriors of the horizon, 
as in the case of the stationary black holes. 
The region $R$ consists of two connected region and $(U_b,v_b)$ and $(U'_b, v'_b)$ 
are the boundaries of the region $R$ in the right and left exteriors, respectively. 
They are related to each other as 
\begin{align}
 U'_b 
 &\simeq 
 v_b - v_0 + V_0 \ , 
\\
 v'_b 
 &\simeq 
 U_b - V_0 + v_0 \ . 
\end{align}
Then, we find 
\begin{equation}
 \frac{c}{6} 
 \log\left|
  \left( U_b - U'_b \right) \left( v_b - v'_b \right)
 \right| 
 \simeq 
 \frac{c}{3} 
 \log\left|
  v_b - v_0 + V_0  
 \right| \ , 
 \label{log-2sided}
\end{equation}
at sufficiently late time. 
It is convenient to define 
\begin{equation}
 V_0 = e^{\kappa(v_0) v_0} \ , 
\end{equation}
and then, the $(U,V)$-coordinates in the stationary spacetime for $v<v_0$ 
is the same to the coordinates in the previous section. 
If $v_0$ is also very large and comparable to $v_b-v_0$, 
\eqref{log-2sided} is further approximated as 
\begin{equation}
 \frac{c}{6} 
 \log\left|
  \left( U_b - U'_b \right) \left( v_b - v'_b \right)
 \right| 
 \simeq 
 \frac{c}{3} 
 \log\left|
  V_0  
 \right| 
 = \frac{c}{3}\, \kappa(v_0)\, v_0 \ . 
\end{equation}
The entanglement entropy is calculated as 
\begin{equation}
 S \simeq 
 \frac{c}{3}\, \kappa(v_0)\, v_0 + \frac{c}{6}\int_{v_0}^{v_b} dv\, \kappa(v) \ , 
\label{NoIdyn}
\end{equation}
where the first term gives the entanglement entropy of the radiation 
during the vacuum state is the Hartle-Hawking vacuum, 
and the second term is the entanglement entropy of Hawking radiation in the Unruh vacuum. 

The entanglement entropy of radiation in the Hartle-Hawking vacuum increases 
twice as fast as that in the Unruh vacuum, since there is both incoming and 
outgoing radiations in the Hartle-Hawking vacuum while the Unruh vacuum has 
only the outgoing Hawking radiation. 
In this paper, we are interested in the $v_b$ dependent part of the entanglement entropy, 
and hence, we consider only the second term in \eqref{NoIdyn}, 
which grows twice as fast as \eqref{NoIeva} since 
\eqref{NoIdyn} is the entanglement entropy of Hawking radiation 
in both two exteriors of the horizon whereas \eqref{NoIeva} 
is for black holes which has only one exterior.


\subsection{Islands in evaporating black holes}\label{ssec:IslandDyn}

Now, we calculate the entanglement entropy of the Hawking radiation 
in the configuration with the island. 
In the case of two sided spacetime, 
the island lies between the quantum extremal surfaces in the right and left exteriors. 
In the case of the one-sided spacetime, 
the island extends to the origin and then is connected to the mirror image. 
In both cases, two boundaries of the island, or equivalently, quantum extremal surfaces 
are sufficiently separated, and it is sufficient to consider only 
one side of the horizon as in the case of stationary black holes. 
Then, the entanglement entropy is given by 
\begin{equation}
 S 
 = 
 S_\text{grav}(U_a,v_a) 
 + \frac{c}{6} \log\left|(U_a-U_b)(v_a-v_b)\right| 
 + \frac{c}{6} \rho(U_a,v_a) 
 + \frac{c}{6} \rho(U_b,v_b) \ . 
\end{equation}
The expression above contains contributions in one side of the horizon. 
Hence, the entanglement entropy in the case of 
one-sided black holes is simply given by the expression above, 
but the entanglement entropy in the case of two-sided black holes is twice of it. 
In the case of the Einstein gravity, 
the gravitational entropy $S_\text{grav}$ is given by 
the area at the boundary of the island, 
\begin{equation}
 S_\text{grav}(U_a,v_a) 
 = 
 \frac{r_h^{D-2}(v_a) \Omega_{D-2}}{4 G_N} 
 \left(1 + (D-2)\frac{r_1(v_a)}{r_h(v_a)} U_a \exp\left[\int_{v_0}^{v_a}dv\,\kappa(v)\right] + \cdots \right) \ . 
\end{equation}
For evaporating black holes, in general, $S_\text{grav}$ takes the following form, 
\begin{equation}
 S_\text{grav}(U_a,v_a) 
 = 
 A(v_a) - B(v_a) U_a \exp\left[\int_{v_0}^{v_a}dv\,\kappa(v)\right] + \cdots \ , 
\end{equation}
where $A(v)$ and $B(v)$ have slow time-dependence. 
The first term $A(v)$ is nothing but the standard black hole entropy $S_\text{BH}$, 
whose change with time is given by \eqref{dSBH}. 
It would be convenient to separate the $v$-dependence of $A$ as 
\begin{equation}
 A(v) \simeq A(v_0) 
 - \frac{c}{24} \int_{v_0}^{v} dv'\kappa(v') \ . 
\end{equation}
As in the case of stationary black hole, 
the second term in \eqref{rho-dyn} is negligible. 
Then, the entanglement entropy is expressed as 
\begin{align}
 S 
 &\simeq 
 A(v_0) + B(v_a) U_a \exp\left[\int_{v_0}^{v_a}dv\,\kappa(v)\right] 
 + \frac{c}{6} \log\left|(U_a-U_b)(v_a-v_b)\right| 
\notag\\
&\quad
 + \frac{c}{24} \int_{v_0}^{v_a} dv\,\kappa(v) 
 + \frac{c}{12} \int_{v_0}^{v_b} dv\,\kappa(v) \ . 
\end{align}

The position of the boundary of the island can be determined 
by the condition of the quantum extremal surface, which is given by 
\begin{align}
 0 
 = 
 \frac{\partial S}{\partial U_a} 
 &=  
 - B(v_a) \exp\left[\int_{v_0}^{v_a}dv\,\kappa(v)\right] + \frac{c}{6} \frac{1}{U_a - U_b} \ ,  
\\
 0 
 = 
 \frac{\partial S}{\partial v_a} 
 &=  
 - B(v_a) \kappa(v_a) U_a \exp\left[\int_{v_0}^{v_a}dv\,\kappa(v)\right] 
 + \frac{c}{6} \frac{1}{v_a - v_b} + \frac{c \kappa(v_a)}{24} \ . 
\end{align}
These conditions have two solutions: 
\begin{align}
 U_a 
 &\simeq 
 - \frac{1}{3} U_b \ , 
&
 \exp\left[\int_{v_0}^{v_a}dv\,\kappa(v)\right] 
 &\simeq 
 - \frac{c}{8B U_b} \ , 
\label{sol-eva}
\end{align}
and 
\begin{align}
 U_a 
 &\simeq 
 U_b + \frac{c}{6 B} \exp\left[- \int_{v_0}^{v_b} dv\,\kappa(v)\right] \ , 
&
 v_a 
 &\simeq 
 v_b + \frac{c}{6 \kappa B U_b} \exp\left[- \int_{v_0}^{v_b} dv\,\kappa(v)\right] \ . 
\label{fake-eva}
\end{align}
The entanglement entropy is calculated as 
\begin{equation}
 S \simeq A(v_0) + \frac{c}{8} \log\left|U_b\right| 
 + \frac{c}{12} \int_{v_0}^{v_b} dv\,\kappa(v) 
\label{S-eva}
\end{equation}
for \eqref{sol-eva}, and 
\begin{equation}
 S \simeq A(v_0) - B U_b \exp\left[\int_{v_0}^{v_b} dv\,\kappa(v)\right] 
 - \frac{c}{6} \log\left|U_b\right| 
 - \frac{5c}{24} \int_{v_0}^{v_b} dv\,\kappa(v) 
\label{S-f-eva}
\end{equation}
for \eqref{fake-eva}. 
When the boundary of the region $R$ is located near the horizon 
but at a distance much larger than the Planck length, namely 
\begin{equation}
 \frac{r_h^2}{A} \ll \left|U_b V_b\right| \ll r_h^2 \ , 
\end{equation}
we have 
\begin{equation}
 \log\left|U_b\right| \simeq - \int_{v_0}^{v_b} dv\,\kappa(v) + \mathcal O(B^{0}) \ . 
\label{logU}
\end{equation}
Since \eqref{S-eva} is smaller than \eqref{S-f-eva}, 
the position of the quantum extremal surface in the dominant saddle point 
is given by \eqref{sol-eva}, and hence, the entanglement entropy is \eqref{S-eva}. 
By using \eqref{logU}, the entanglement entropy is expressed 
in terms of the advanced time $v$ as 
\begin{equation}
 S \simeq A(v_0) - \frac{c}{24} \int_{v_0}^{v_b} dv\,\kappa(v) \simeq A(v_b) \ . 
\label{Ieva}
\end{equation}
Thus, the entanglement entropy is approximately 
the same to the black hole entropy. 

The quantum extremal surface is located inside the horizon since $U_b > 0$. 
The position \eqref{sol-eva} at the leading order is independent of 
details of the black hole geometry or matter fields. 
It should be noted that the difference between $v_a$ and $v_b$ 
is of the same order to the scrambling time, 
\begin{equation}
 v_b - v_a \simeq \kappa^{-1} \log B \ , 
\end{equation}
and the change of the black hole entropy within 
a period comparable with the scrambling time 
has been ignored in \eqref{Ieva} as a higher order correction.


\subsection{Black holes with conserved charges}\label{ssec:chargedBH}

In this section, we have studied the entanglement entropy of 
Hawking radiation around an evaporating black hole. 
The entanglement entropy is different from but related to the thermal entropy. 
The thermal entropy of the black hole is given by the Wald entropy $S_\text{grav}$, 
and the thermal entropy of the Hawking radiation approximately equals to 
the conformal factor in the Eddington-Finkelstein coordinates $S_\text{rad} \simeq \frac{c}{6}\rho$. 
These two entropies are related to each other 
by the first law of thermodynamics \eqref{thermo}. 
The black hole may have a conserved charge, 
but the relation \eqref{thermo} contains only the change 
of the black hole entropy and the energy taken away by radiation. 
The thermodynamic relation including the charges and angular momenta 
is given by 
\begin{equation}
 \delta M = \frac{\kappa}{2\pi} \delta S_\text{BH} + \Omega_H \delta J + \mu \delta Q \ , 
 \label{thermo0}
\end{equation}
where $J$ is the angular momentum, $Q$ is a charge, $\Omega_H$ is the velocity of the horizon, 
and $\mu$ is the chemical potential associated to the charge $Q$. 
Although the relation \eqref{thermo} is written only in terms of the energy and entropy, 
it already contains effects of conserved charges above. 

The variation of the energy $\delta E$ is not equal to 
the variation of the black hole mass $\delta M$, 
since the energy $E$ in \eqref{thermo} is measured near the horizon. 
For rotating black holes, the time $t$ and angular coordinate $\varphi$ in the asymptotic region 
is related to the comoving coordinates to the horizon as 
\begin{align}
 t_\text{com} &= t_\text{asym} \ , 
 &
 \varphi_\text{com} &= \varphi_\text{asym} + \Omega_H t_\text{asym} \ , 
\end{align}
Then, the energy of matter fields near the horizon is related to those in the asymptotic region as 
\begin{align}
 \omega_\text{com} &= \omega_\text{asym} - \Omega_H m_\text{asym} \ , 
 &
 m_\text{com} &= m_\text{asym} \ , 
\end{align}
where $\omega$ and $m$ are the energy (frequency) and the angular momentum of the field, respectively. 
Thus, the change of the energy near the horizon and 
the mass of the black hole is related to each other as 
\begin{equation}
 \delta E = \delta M - \Omega_H \delta J \ . 
\end{equation}

In a similar fashion, the variation of the charge is also included in $\delta E$. 
The chemical potential $\mu$ is nothing but the difference of the potential energy 
between the horizon and spatial infinity. 
A charged particle with the energy $\delta E$ near the horizon 
has the energy 
\begin{equation}
 \delta E + \mu \delta Q \ , 
\end{equation}
when it reach the spatial infinity, and hence, the black hole (without angular momenta) loses 
the mass $\delta M = \delta E + \mu \delta Q$ by the emission of this particle. 

Thus, the variations of charges and angular momenta are included in 
the variation of the energy near the horizon, 
\begin{equation}
 \delta E = \delta M - \Omega_H \delta J - \mu \delta Q \ . 
\end{equation}
The first law of the black hole \eqref{thermo0} is equivalent to 
the first law near the horizon \eqref{thermo}, 
and hence, the change of the black hole entropy is simply given by \eqref{dSBH} 
even for black holes with charges and angular momenta. 
In fact, the energy of matter near the horizon is associated to the horizon generator, 
which is nothing but the Killing vector associated to the Wald entropy. 
Thus, the change of the black hole entropy is 
given by the energy of Hawking radiation.


\section{Conclusion}\label{sec:Conclusion}

Islands appears when we use the replica trick in theories with gravity 
and plays an important role in reproducing the Page curve. 
It is expected that the entanglement entropy of the Hawking radiation 
approximately agrees with the black hole entropy 
independent of details of the model. 
In this paper, we have studied the entanglement entropy of the Hawking radiation around general black holes. 
We have seen explicitly that calculations 
by using the island rule actually do not depend on details of the model. 

In the case of stationary black holes, the boundaries of islands, or equivalently, 
quantum extremal surfaces are shifted from the classical extremal surface 
by a small correction by the matter part of the entanglement entropy. 
The gravity part of the entanglement entropy is extremized at the classical extremal surface, 
which is nothing but the bifurcation surface of the horizon, 
but the island disappear for the classical extremal surface since 
both of two boundaries are located on the same place. 
By taking effects of the matter part into account, 
the extremal surface is moved outward from the horizon. 
Then, two boundaries of the island is separated from each other, 
and the island has a large volume. 
The quantum extremal surfaces are still located very near the horizon, 
and contributions from the quantum extremal surfaces in the gravity part 
give the dominant contribution to the entanglement entropy. 
Thus, the entanglement entropy of Hawking radiation 
is approximately given by the black hole entropy. 
The entanglement entropy can be calculated by using two important behaviors:  
(1) the gravity part is extremized at the bifurcation surface, 
and (2) the matter part is monotonically increases as the distance between 
the region of Hawking radiation and the island increases. 

In the case of evaporating black holes, 
the incoming negative energy near the horizon 
plays an important role in determining position of the quantum extremal surface. 
In the presence of the incoming negative energy, 
the matter part of the entanglement entropy has additional finite local terms 
which are given in terms of the conformal factor $\rho$ of the two-dimensional metric 
in the temporal and radial directions. 
The thermal behavior of the entanglement entropy of Hawking radiation 
at early times is reproduced by this local term. 
This local term pushes the quantum extremal surface into the horizon. 
At the same time, the gravity part of the entanglement entropy also 
plays an important role. 
In the case of the Einstein gravity, the gravity part is given by the area of the horizon, 
which is determined by the overall factor $r^2$ of the metric in the angular directions. 
The decrease in the horizon area pulls the quantum extremal surface outward from the horizon. 
Thus, the position of the quantum extremal surface 
is determined by the relation between these two terms. 
The quantum extremal surface is inside the horizon if the matter part is larger and 
outside the horizon if the change of the area term is larger. 
The relation is universally determined by the first law of the thermodynamics, 
which relates the change of the black hole entropy with emission of Hawking radiation. 
The local term in the matter part is always twice of the change of the black hole entropy, 
and hence, the quantum extremal surface is always inside the horizon. 

Thus, the position of the quantum extremal surface is independent of details of the model. 
The quantum extremal surface is located outside the horizon for stationary black holes 
and inside the horizon for evaporating black holes. 
The difference between stationary cases and dynamical cases mainly 
comes from the different vacuum states: the Hartle-Hawking state for stationary black holes 
and the Unruh vacuum for evaporating black holes, 
and secondly comes from the change of the position of the horizon. 
Once the vacuum state is given, the position of 
the quantum extremal surface is almost fixed. 
In other word, the region in which Hawking radiation has the information 
is universally determined by the vacuum state, 
which is nothing but the state of Hawking radiation.


\section*{Acknowledgments}

This work is supported in part by JSPS KAKENHI Grant No.~JP20K03930 and JP21H05186.

\appendix


\section*{Appendix}


\section{Vacuum states and associated coordinates}

Here, we discuss the relation between coordinates and vacuum states. 
Although physics does not depend on the choice of the coordinate system, 
physical states are not invariant under the coordinate transformation. 
Correlation functions even do not have covariant expressions 
and can be expressed in a simple form in some specific coordinates. 
The vacuum state is the state with minimum energy, 
which depends on the definition of the energy, or equivalently time. 
Thus, even vacuum states are not invariant under the coordinate transformation. 
Vacuum states are not unique, and each of them is associated to a time coordinate. 

For simplicity, we consider a two-dimensional massless free scalar field $\phi$. 
The action is given by 
\begin{equation}
 S = - \frac{1}{2}\int d^2x \sqrt{-g} g^{\mu\nu} \partial_\mu \phi \partial_\nu \phi 
 = - \frac{1}{2}\int du\,dv\,\partial_u \phi \partial_v \phi \ , 
\end{equation}
where $u$ and $v$ are the retarded and advanced time coordinates. 
The equation of motion is expressed as 
\begin{equation}
 0 = \partial_\mu \sqrt{-g}\,g^{\mu\nu} \partial_\nu\phi = \partial_u \partial_v \phi \ , 
\end{equation}
and the solution is separated into incoming modes $\phi_+$ and outgoing modes $\phi_-$ as 
\begin{equation}
 \phi(x) = \phi_+(v) + \phi_-(u) \ , 
\end{equation}
where $\phi_+(v)$ and $\phi_-(u)$ are arbitrary functions of 
the advanced time $v$ and the retarded time $u$, respectively. 
They can be expanded in the Fourier modes, and then, 
the solution $\phi$ is expressed as 
\begin{equation}
 \phi(x) 
 = 
 \int_0^\infty \frac{ d \omega}{2\pi} \frac{1}{\sqrt{2\omega}} 
 \left[a_\omega e^{-i\omega v} + a_\omega^\dag e^{i\omega v}
 + b_\omega e^{-i\omega u} + b_\omega^\dag e^{i\omega u}\right] \ . 
\end{equation}
The coefficients $a_\omega$, $a_\omega^\dag$, $b_\omega$ and $b_\omega^\dag$ are 
annihilation and creation operators. 
The annihilation operators, $a_\omega$ and $b_\omega$ are coefficients of 
positive frequency modes, $e^{-i \omega v}$ and $e^{-i \omega u}$, respectively, 
and the creation operators, $a_\omega^\dag$ and $b_\omega^\dag$ are 
coefficients of negative frequency modes. 
The vacuum state $|0\rangle$ is defined as the state 
which is annihilated by the annihilation operators $a_\omega$ and $b_\omega$, 
\begin{align}
 a_\omega |0\rangle &= 0 \ , 
 &
 b_\omega |0\rangle &= 0 \ . 
\end{align}
Here, the mode expansion is defined in terms of the coordinates $u$ and $v$, 
and hence, the vacuum state $|0\rangle$ is also associated to these coordinates. 
For example, the two-point correlation function is calculated as 
\begin{equation}
 \langle 0 | \phi(x) \phi(x') |0\rangle 
 = - \frac{1}{4\pi}\log \left|(u-u')(v-v')\right| \ , 
\label{Green0}
\end{equation}
but the expression is not covariant. 
The distance is not written in terms of the proper distance 
but simply the difference of the coordinates $u$ and $v$ between two points.  

Since the solution is given in terms of arbitrary functions of $u$ and $v$, 
we can consider the mode expansion in any other null coordinates. 
We define another pair of null coordinates as $U=U(u)$ and $V=V(v)$, 
and expand $\phi$ in the Fourier modes in these coordinates as 
\begin{equation}
 \phi(x) 
 = 
 \int_0^\infty \frac{ d \omega}{2\pi} \frac{1}{\sqrt{2\omega}} 
 \left[A_\omega e^{-i\omega V} + A_\omega^\dag e^{i\omega V}
 + B_\omega e^{-i\omega U} + B_\omega^\dag e^{i\omega U}\right] \ , 
\end{equation}
and then, another vacuum state $|\Omega\rangle$ is defined by using 
the annihilation operators $A_\omega$ and $B_\omega$ above, 
\begin{align}
 A_\omega |\Omega\rangle &= 0 \ , 
 &
 B_\omega |\Omega\rangle &= 0 \ . 
\end{align}
The vacuum state $|\Omega\rangle$ is in general different from $|0\rangle$ 
since annihilation operators defined in different coordinates are different. 
The annihilation operator $a_\omega$ is related to $A_\omega$ and $A_\omega^\dag$ as 
\begin{equation}
 a_\omega 
 = 
 \int d \omega' 
 \left(\mathcal A_{\omega\omega'} A_{\omega'} 
 + \mathcal A_{\omega,-\omega'} A_{\omega'}^\dag \right) \ , 
\end{equation}
where 
\begin{equation}
 \mathcal A_{\omega\omega'} = \sqrt{ \frac{\omega}{\omega'}} \int dv\, e^{i \omega v}e^{-i \omega' V(v)} \ . 
\end{equation}
Then, the vacuum state $|\Omega\rangle$ is not annihilated by $a_\omega$ or $b_\omega$ since 
\begin{align}
 a_\omega |\Omega\rangle 
 = 
 \int d \omega' 
 \mathcal A_{\omega,-\omega'} A_{\omega'}^\dag |\Omega\rangle 
 \neq 0 \ ,  
\end{align}
and similarly $b_\omega |\Omega\rangle \neq 0$. 
This implies that the state $|\Omega\rangle$ is viewed as an excited state 
for an observer in the $(u,v)$-coordinates. 
Conversely, the vacuum state $|0\rangle$ is not annihilated by 
annihilation operators $A_\omega$ and $B_\omega$, and 
is an excited state from the viewpoint of $(U,V)$-coordinates. 

The two-point correlation function in the vacuum state $|\Omega\rangle$ 
is calculated in a similar fashion to \eqref{Green0} as 
\begin{align}
 \langle \Omega | \phi(x) \phi(x') |\Omega \rangle 
 = - \frac{1}{4\pi}\log \left|(U-U')(V-V')\right| \ , 
 \label{GreenO}
\end{align}
which is obviously different from \eqref{Green0}. 

The expectation value of the energy-momentum tensor depends on the physical state 
and takes different values in different vacua. 
We first consider the energy-momentum tensor in the flat spacetime, 
\begin{equation}
 ds^2 = - du\ dv \ . 
 \label{flat}
\end{equation}
The energy-momentum tensor is a composite operator and 
defined by taking the normal ordering. 
The expectation value of $T_{uu}$ is given by 
\begin{equation}
 \langle T_{uu}\rangle 
 = 
 \lim_{x'\to x} \langle \psi | : \partial_u\phi(x) \, \partial_{u'}\phi(x') : |\psi \rangle \ , 
\end{equation}
and similarly for other components. 
The normal ordering is defined by using the annihilation and creation operators 
in the coordinates of the flat spacetime \eqref{flat}, namely the $(u,v)$-coordinates. 
The expectation value $\langle T_{uu} \rangle$ vanishes 
in the vacuum state $|0\rangle$ by definition. 
In fact, it is straightforward to see that the normal ordering is 
equivalent to subtracting the two-point function in the vacuum state $|0\rangle$, 
\begin{equation}
 : \phi(x) \, \phi(x') : \ 
 = 
 \phi(x)\,\phi(x') 
 + \frac{1}{4\pi}\log \left|(u-u')(v-v')\right| \ . 
\end{equation}
If the state is the vacuum state, $|\phi\rangle = |\Omega\rangle$, 
the expectation value is calculated as 
\begin{align}
 \langle \Omega| T_{uu} |\Omega \rangle
 &= - \frac{1}{4\pi} \lim_{u'\to u} 
 \partial_u \partial_{u'} \log\left|\frac{U(u)-U(u')}{u-u'}\right|  
 \notag\\
 &= - \frac{1}{24\pi} \left\{U,u\right\} \ , 
\end{align}
where $\{f,x\}$ is the Schwarzian derivative. 

However, the normal ordering of the annihilation and creation operators 
obviously depends on the coordinates as the operators are associated to a time coordinate. 
In the curved spacetime, we need to the energy-momentum tensor 
by removing the divergence in a covariant manner. 
Hence, we define the energy-momentum tensor by subtracting the proper geodesic distance $\Delta s$, 
\begin{equation}
 \langle T_{uu}\rangle 
 = 
 \lim_{x'\to x} \partial_u \partial_{u'} 
 \left(
  \langle\phi(x)\,\phi(x') \rangle 
  + \frac{1}{4\pi} \log \left|\Delta s(x,x')\right|^2 
 \right) \ . 
\end{equation}
In calculating the expression above, we can first take the limit $v' \to v$ before the differentiation, 
and hence, the $v$-dependence of the metric can be ignored. 
The metric on the null surface with $v=\text{const.}$ can be written in terms of an affine parameter $U$ as 
\begin{equation}
 ds^2 = - C(u,v) du\,dv = - dU\,dv \ , 
 \label{curved}
\end{equation}
and then, the geodesic distance $\Delta s$ is given by 
\begin{equation}
 \left|\Delta s\right|^2 
 = 
 \left|(\Delta U)(\Delta v)\right| \ , 
\end{equation}
where $\Delta U$ is calculated as 
\begin{equation}
 \Delta U
 = 
 \int_u^{u'} du\, C(u,v) \ . 
\end{equation}
Thus, the expectation value of $T_{uu}$ is given by 
\begin{equation}
 \langle T_{uu} \rangle 
 = 
 \lim_{x'\to x} 
 \left(
  \partial_u\phi(x) \, \partial_{u'}\phi(x')
  + \frac{1}{4\pi} \frac{1}{|u-u'|^2} 
 \right) 
 - \frac{1}{12\pi} C^{1/2} \partial_u^2 C^{-1/2} 
 \ . 
 \label{T-reg}
\end{equation}
It should be noted that the combination 
\begin{equation}
 \lim_{x'\to x} \left(\frac{1}{4\pi}
  \frac{1}{|u-u'|^2} \right) 
 - \frac{1}{12\pi} C^{1/2} \partial_u^2 C^{-1/2} 
\end{equation}
is covariant under the coordinate transformation, 
and the regularized energy-momentum tensor \eqref{T-reg} 
has no artifact of the regularization. 
Since the combination 
\begin{equation}
 \langle\phi(x)\,\phi(x')\rangle 
 + \frac{1}{4\pi} \log\left|(u-u')(v-v')\right|
\end{equation}
is regular and can be separated into incoming modes and outgoing modes, 
the expectation value can always be written in the following form:%
\footnote{%
Here, we consider the case of a single free massless scalar field, and hence, $c=1$. 
} 
\begin{equation}
 \langle T_{uu} \rangle 
 = 
 - \frac{1}{12\pi} C^{1/2} \partial_u^2 C^{-1/2} + F(u)
 \ , 
\end{equation}
where $F(u)$ is a regular function of $u$. 
For the vacuum state $|0\rangle$ which is associated to the coordinates in \eqref{curved}, 
the expectation value of $T_{uu}$ becomes 
\begin{equation}
 \langle T_{uu} \rangle 
 = 
 - \frac{1}{12\pi} C^{1/2} \partial_u^2 C^{-1/2} \ , 
\end{equation}
and hence, $F(u)$ vanishes. 

Now, it is clear how the entanglement entropy depends on the vacuum state. 
By using the replica trick, the entanglement entropy $S$ is expressed as 
\begin{equation}
 S = \lim_{n\to 1} \frac{1}{1-n} \log \frac{Z_n}{Z_1^n} \ , 
\end{equation}
where $Z_n$ is the partition function on $n$-sheeted Riemann surface, 
which can be calculated by inserting twist operators at the branch points. 
For free massless fermions in two dimensions, 
the partition function $Z_n$ is given by using the bosonization as \cite{Casini:2005rm}, 
\begin{equation}
 Z_n 
 = 
 \prod_{k} 
 \left\langle \prod_i e^{i\sqrt{4\pi}\frac{k}{n}\phi(x_i)} 
 e^{-i\sqrt{4\pi}\frac{k}{n}\phi(x'_i)} \right\rangle \ , 
\end{equation}
where $e^{\pm i\sqrt{4\pi}\frac{k}{n}\phi}$ are the twist operators on the $k$-th sheet, 
and the branch points are located at $x_i$ and $x'_i$. 
Thus, the entanglement entropy is given in terms of the correlation function of the bosonized field, 
\begin{equation}
 S 
 = 
 - \frac{2\pi}{3} \sum_{i,j}\langle\phi(x_i)\,\phi(x'_j)\rangle
 + \frac{\pi}{3} \sum_{i,j}\langle\phi(x_i)\,\phi(x_j)\rangle
 + \frac{\pi}{3} \sum_{i,j}\langle\phi(x'_i)\,\phi(x'_j)\rangle \ , 
\end{equation}
which depends on the physical state. 
If the state is the vacuum state $|0\rangle$, 
the correlation function is expressed in terms of the associated coordinates $u$ and $v$, 
\begin{equation}
 \langle 0|\phi(x)\,\phi(x')|0\rangle 
 = - \frac{1}{4\pi}\log\left|(u-u')(v-v')\right| \ . 
\end{equation}

The entanglement entropy has divergence but the regularization should be given 
in a similar fashion to the energy-momentum tensor. 
The geodesic distance is approximated simply as 
\begin{equation}
 (\Delta s)^2 \simeq - C (\Delta u)(\Delta v) \ , 
\end{equation}
in $x'\to x$, since we do not need to differentiate it in this case. 
Then, the entanglement entropy can be separated into the divergent part and regular part, 
\begin{align}
 S 
 &= 
 S_\text{reg} 
 - \frac{k}{3}\log \epsilon
\\
 S_\text{reg}
 &= 
 - \frac{2\pi}{3} \sum_{i,j}\langle\phi(x_i)\,\phi(x'_j)\rangle
 + \frac{2\pi}{3} \sum_{i<j}\langle\phi(x_i)\,\phi(x_j)\rangle
 + \frac{2\pi}{3} \sum_{i<j}\langle\phi(x'_i)\,\phi(x'_j)\rangle 
\notag\\
&\quad
 + \frac{\pi}{3} \sum_i \langle\phi(x_i)\,\phi(x_i)\rangle_\text{reg} 
 + \frac{\pi}{3} \sum_i \langle\phi(x'_i)\,\phi(x'_i)\rangle_\text{reg} 
 \ , 
\end{align}
where $k$ is the number of the branch cuts. The regular part of the two-point function is defined by 
\begin{equation}
 \langle\phi(x)\,\phi(y) \rangle_\text{reg} 
 = 
 \langle\phi(x)\,\phi(y) \rangle 
 + \frac{1}{4\pi} \log \left|\Delta s(x,y)\right|^2 \ , 
\end{equation}
and behaves in the vacuum state as 
\begin{equation}
 \langle 0| \phi(x)\,\phi(x) |0\rangle_\text{reg} 
 = \frac{1}{4\pi}\log C(x) \ , 
\end{equation}
where $C$ is the conformal factor of the metric in coordinates associated to the vacuum state.


\end{document}